\documentclass[10pt,a4paper]{article}
\pdfoutput=1
\usepackage{times}
\usepackage{a4wide}
\usepackage{graphicx}
\usepackage{amsfonts}
\usepackage{amssymb}
\usepackage{amsmath}
\usepackage{ifpdf}
\ifpdf

\usepackage[sort&compress,comma]{natbib}
\bibpunct{[}{]}{,}{n}{}{,}

\usepackage[pdftex,unicode,implicit]{hyperref}
\hypersetup{%
  pdftitle    = {On hvLif-like solutions in gauged Supergravity},
  pdfkeywords = {String theory, gauged supergravity, Lifshitz, Hyperscaling violation, AdS/CMP},
  pdfauthor   = {P.~Bueno, W.~Chemissany, C.~S.~Shahbazi},
  plainpages  = true,
  colorlinks  = true,
  citecolor   = blue,
  urlcolor    = red,
  linkcolor   = black
}

\else
  \usepackage[dvips]{graphicx}
  \usepackage[unicode,implicit]{hyperref}

\fi
\makeatletter
\@addtoreset{equation}{section}
\makeatother

\begin{document}

\begin{flushright}
\small
IFT-UAM/CSIC-12-121\\
\texttt{arXiv:1212.4826 [hep-th]}\\
January 11\textsuperscript{th}, 2013\\
\normalsize
\end{flushright}

\vspace{1cm}

\begin{center}

{\LARGE {  On \emph{hvLif}-like solutions in gauged Supergravity}}

\vspace{.7cm}

\renewcommand{\thefootnote}{\alph{footnote}}
{\sl\large P.~Bueno$^{\heartsuit} $}
\footnote{E-mail: {\tt pab.bueno [at] estudiante.uam.es}},
{\sl\large W.~Chemissany$^{\clubsuit, \diamondsuit}$}
\footnote{E-mail: {\tt chemissany.wissam [at] gmail.com}}
\footnote{Riemann Fellow at the Riemann Center for Geometry and Physics, Leibniz Universit\"at Hannover
Germany.},
{\sl\large and C.~S.~Shahbazi$^{\heartsuit}$}
\footnote{E-mail: {\tt Carlos.Shabazi [at] uam.es}}
\renewcommand{\thefootnote}{\arabic{footnote}}

\vspace{.4cm}

${}^{\heartsuit}${\it Instituto de F\'{\i}sica Te\'orica UAM/CSIC\\
C/ Nicol\'as Cabrera, 13--15,  C.U.~Cantoblanco, 28049 Madrid, Spain}\\

\vspace{.2cm}
${}^{\diamondsuit}${\it Department of Physics and Astronomy, University of Waterloo, Waterloo, Ontario, Canada, N2L 3G1}\\

\vspace{.2cm}

${}^{\clubsuit}${\it Institut f\"ur Theoretische Physik \& Riemann Center for Geometry and Physics, Leibniz Universit\"at Hannover, Appelstra\ss e 2, 30167 Hannover, Germany}\\

\vspace{2.5cm}

{\bf Abstract}

\begin{quotation}

  {\small
We perform a thorough investigation of Lifshitz-like metrics with hyperscaling violation (\emph{hvLif}) in four-dimensional theories of gravity coupled to an arbitrary number of scalars and vector fields, obtaining new solutions, electric, magnetic and dyonic, that include the known ones as particular cases. After establishing some general results on the properties of purely \emph{hvLif} solutions, we apply the previous formalism to the case of $\mathcal{N}=2,~d=4$ Supergravity in the presence of Fayet-Iliopoulos terms, obtaining particular solutions to the $t^3$-model, and explicitly embedding some of them in Type-IIB String Theory.
}

\end{quotation}

\end{center}

\setcounter{footnote}{0}

\newpage
\pagestyle{plain}

\tableofcontents


\section*{Introduction}
Gauge/gravity duality  has been shown to be an instrumental tool to study strongly coupled systems near critical points where the system displays a scaling symmetry. Generically, conformal field theories provide consistent descriptions of certain physical systems near critical points.  In the gauge/gravity  avatar this means that the gravitational theory is living on a background which is asymptotically locally anti De Sitter (\emph{aDS}). On the other hand, in many physical systems critical points are dictated by dynamical scalings in which, even though the system exhibits a scaling symmetry, space and time scale differently under this symmetry. A prototype example of such critical points is a hyperscaling violating Lifshitz  fixed point where the system is spatially isotropic and scale covariant, though there is an anisotropic scaling in the time direction characterized by a dynamical exponent, $z$, and hyperscaling violation characterized by the exponent $\theta.$ More precisely the system is covariant under the following scale symmetry \cite{Huijse:2011ef,Iizuka:2011hg,Shaghoulian:2011aa,Iizuka:2012pn,Ogawa:2011bz,Edalati:2012tc,Kim:2012pd,Alishahiha:2012qu,Gath:2012pg,Gouteraux:2012yr,Alishahiha:2012cm,Narayan:2012hk,Dey:2012tg,Dey:2012rs,Charmousis:2012dw,Ammon:2012je,Bhattacharya:2012zu,Dey:2012fi,Kim:2012nb,Perlmutter:2012he,Gouteraux:2011qh,Sadeghi:2012vv,Donos:2012yu}
 \begin{equation} x_{i} \to \lambda x_{i},\quad t \to \lambda^{z} t, \quad r \to \lambda r, \quad ds^{2}_{d+2} \to \lambda ^{2\theta/d} ds_{d+2},\end{equation}
 where $\lambda$ is a dimensionless parameter and $d$ is the number of spatial dimensions on which the dual theory lives $(i=1,\cdots, d).$ The value $\theta=0$ corresponds to the standard scale-invariant theories dual to Lifshitz metric \cite{Copsey:2010ya,Copsey:2012gw,Bao:2012yt,Copsey:2011ek,Horowitz:2011gh,Balasubramanian:2011ua}. The values $z=1$ and $\theta=0$ correspond to conformally-invariant theories dual to gravity theories on an \emph{aDS} background. For other values of $z$ and $\theta,$ the $d+2$-dimensional gravitational backgrounds are supported by metrics of the form
 \begin{equation}\label{hvLif} d^{2}_{d+2}=\ell^{2} r^{-2(d-\theta)/d}\left(r^{-2(z-1)} dt^{2}-dr^{2}-dx_{i} dx_{i}\right),\end{equation}
 where $\ell$ is the Lifshitz radius . As usual, we will refer to these metrics as hyperscaling-violating Lifshitz (\emph{hvLif}) metrics.  The Lifshitz-type  spacetimes  are known to be singular in the IR. They suffer from a null singularity with diverging tidal forces \cite{Kachru:2008yh,Taylor:2008tg,Azeyanagi:2009pr,Balasubramanian:2010uk,Donos:2010tu,Gregory:2010gx,Narayan:2011az,Chemissany:2011mb,Chemissany:2012du,Danielsson:2009gi,Bertoldi:2009vn,Balasubramanian:2009rx,Tarrio:2011de}. For holographic related applications, see \cite{Hartnoll:2009sz,Herzog:2009xv,McGreevy:2009xe,Sachdev:2011wg,Hartnoll:2009ns,Donos:2010ax,Halmagyi:2011xh,Hartnoll:2012wm,Takayanagi:2012kg,Dey:2012hf,Kulaxizi:2012gy,Charmousis:2010zz}.\\

It is interesting to obtain new gravitational solutions that may be used as duals of the corresponding field theories, if any. A first step on this direction, for $\mathcal{N}=2,~ d=4$ ungauged Supergravity was taken in \cite{Bueno:2012sd}, where a complete analysis on the existence of such kind of solutions was performed. In this note we extend the systematic study to a general class of gravity theories coupled to scalars and vectors, up to two derivatives, in the presence of a scalar potential, in principle arbitrary, focusing later on $\mathcal{N}=2,~d=4$ Supergravity in the presence of Fayet-Iliopoulos terms. \\

 The structure of the paper goes as follows: in section 1 we dimensionally reduce the general action of gravity coupled to an arbitrary number of scalars and vectors in the presence of a scalar potential assuming a general static background which naturally fits the anisotropic scaling properties which correspond to \emph{hvLif}-like solutions. In section 2 we adapt the general formalism to the Einstein-Maxwell-Dilaton system. In section 3 we focus on  $\mathcal{N}=2,~d=4$ Supergravity in the presence of Fayet-Iliopoulos terms (which correspond to include a scalar potential in ungauged Supergravity), were we exploit the symplectic structure of the theory in order to obtain further results. We also embed a particular truncation of the $t^3$-model  in Type-IIB String Theory compactified on a Sasaki-Einstein manifold times $S^1$. In section 4 we perform an analysis of the properties of purely \emph{hvLif} solutions for the general class of theories considered. In addition, we provide a general recip to obtain \emph{hvLif}-like solutions of a particular class of Einstein-Maxwell-Dilaton systems, reducing the problem to the resolution of an algebraic equation. We apply the procedure to obtain explicit solutions, some of them embedded in String Theory. In section 5 we conclude.


\section{The general theory}
\label{sec:effectiveaction}

We are interested in Lifshitz-like solutions with hyper-scaling violation (\emph{hvLif}\footnote{We will understand for \emph{hvLif} any non-trivial gravitational solution that presents some kind of Lifshitz limit with hyper-scaling violation. \emph{Purely hvLif} stands for metrics that are exactly Lifshitz with hyper-scaling violation.}) of the four-dimensional action

\begin{equation}
\label{eq:action}
S= \int d^{4}x \sqrt{|g|}\,
\left\{
R  +\mathcal{G}_{ij}\partial_{\mu}\phi^{i}\partial^{\mu}\phi^{j}
+2I_{\Lambda\Sigma}F^{\Lambda}{}_{\mu\nu}F^{\Sigma\, \mu\nu}-2R_{\Lambda\Sigma}F^{\Lambda}{}_{\mu\nu} \star F^{\Sigma\, \mu\nu}
-V(\phi)
\right\}\, ,
\end{equation}

\noindent
that generalizes the action considered in Ref.~\cite{Ferrara:1997tw,Galli:2011fq} by
including a generic scalar potential $V(\phi)$. We will take care of the constraints imposed by  $\mathcal{N}=2$ supersymmetry
on the field content, the kinetic matrices $\left(I_{\Lambda\Sigma}(\phi)<0,~R_{\Lambda\Sigma}(\phi)\right)$, the scalar metric $\mathcal{G}_{ij}(\phi)$ and the scalar potential $V(\phi)$  later on.

The idea now is to dimensionally reduce the action (\ref{eq:action}) using an appropriate ansatz for the metric. Since \emph{hvLif} solutions are in particular static, a first step is to constrain the form of the metric to be

\begin{equation}
\label{eq:metricageneral}
  ds^{2} = e^{2U}dt^{2} - e^{-2U}\gamma_{\underline{m}\,
    \underline{n}}dx^{\underline{m}} dx^{\underline{m}}\, ,~~~\underline{m},\underline{n}=1,\dots ,3 \, ,
\end{equation}

\noindent
A sensible choice for $\gamma$, that fits the anisotropic scaling properties that we look for in a \emph{hvLif} solution, is given by

\begin{equation}
\label{eq:gamma}
\gamma = \gamma_{\underline{m}\, \underline{n}}dx^{\underline{m}} dx^{\underline{m}}
= e^{2W}\left(dr^{2} +\delta_{ab}dx^a dx^b\right) ,~~~ a,b=1,2\, ,
\end{equation}

\noindent
where $e^{W}$ is an undetermined function of the ``radial'' coordinate $r$.
\noindent
We now proceed to dimensionally reduce the lagrangian (\ref{eq:action}) with the choice of metric given by Eqs. (\ref{eq:metricageneral}) and (\ref{eq:gamma}).

Assuming that all the fields are static, only depend on $r$, and following
the same steps as in Refs.~\cite{Ferrara:1997tw,Meessen:2011bd}\footnote{A related procedure, used to obtain non-extremal $aDS_4$ black hole solutions can be found in \cite{Klemm:2012yg} and \cite{Klemm:2012vm}. For related Refs. about solutions in gauged Supergravity see \cite{Barisch:2011ui,BarischDick:2012gj}.}, one arrives to a set of equations of motion for the variables $U(r),~W(r),~\phi^{i}(r)$ that can be derived from the following effective action $\left(^{\prime}=\frac{d}{dr}\right)$

\begin{equation}
\label{eq:effectiveaction}
  \begin{array}{rcl}
S
=
{\displaystyle
\int dr\, e^{W}\,
\left\{
2 U^{\prime\, 2}-2 W^{\prime\, 2}
+ \mathcal{G}_{ij}\phi^{i\,\prime}\phi^{j\,\prime}
-2 e^{2(U-W)}V_{\rm bh}
+e^{-2(U-W)}V
\right\}\, ,
}
\end{array}
\end{equation}

\noindent
if we set the value of the Hamiltonian (which is conserved, due to the lack of
explicit $r$-dependence of the Lagrangian) to zero, that is:

\begin{equation}
\label{eq:hamiltonian}
 2U^{\prime\, 2}-2W^{\prime\, 2}
+ \mathcal{G}_{ij}\phi^{i\, \prime}\phi^{j\, \prime}
+2e^{2(U-W)}V_{\rm bh}
-e^{-2(U-W)}V
=
0\, .
\end{equation}

\noindent
The \textit{black-hole potential} $V_{\rm bh}$ is defined as in the
asymptotically-flat case by \cite{Ferrara:1997tw,Meessen:2011bd,deAntonioMartin:2012bi}

\begin{equation}
\label{eq:Vbh}
V_{\rm bh}(\phi,\mathcal{Q}) \equiv
2 \alpha^{2}
\mathcal{M}_{MN}(\phi) \mathcal{Q}^{M}\mathcal{Q}^{N}\, ,
\end{equation}

\noindent
where $\alpha$ is a normalization constant for the electric and
magnetic charges\footnote{The canonical choice for $d=4$ is $\alpha= \frac{1}{2}$.}

\begin{equation}
\left( \mathcal{Q}^{M} \right)
=
\left(
  \begin{array}{c}
   p^{\Lambda} \\ q_{\Lambda} \\
  \end{array}
\right)\, ,
\end{equation}

\noindent
and where the symmetric matrix $\mathcal{M}_{MN}$ is defined in terms of $I
\equiv (I_{\Lambda\Sigma})$ and $R\equiv (R_{\Lambda\Sigma})$ by

\begin{equation}
\label{M}
\left(\mathcal{M}_{MN} \right)
\equiv
\left(
\begin{array}{cc}
I+RI^{-1}R  & -RI^{-1} \\
& \\
-I^{-1}R & I^{-1} \\
\end{array}
\right)\, .
\end{equation}

\noindent
The one-dimensional effective equations of motion are given by\footnote{The form of the vector fields can be recovered following the dimensional-reduction procedure. The corresponding field strengths $F^{\Lambda}_{\mu\nu}$ are given by

\begin{equation}
F^{\Lambda}_{\underline{m}t} = -\partial_{\underline{m}}\psi^{\Lambda}\, ,\qquad F^{\Lambda}_{\underline{m}\underline{n}} = \frac{e^{-2U}}{\sqrt{|\gamma|}}\epsilon_{\underline{m}\underline{n}\tau}\left[ \left( I^{-1}\right)^{\Lambda\Omega}\partial_{\tau}\chi_{\Omega} - \left( I^{-1} R\right)^{\Lambda}_{\Omega}\partial_{\tau}\psi^{\Omega}\right] \, ,
\end{equation}

\noindent
where $\Psi = (\psi^{\Lambda},\chi_{\Lambda})^{T}$ is a symplectic vector whose components are the time components of the electric $A^{\Lambda}$ and magnetic $A_{\Lambda}$ vector fields. $\Psi$ is given by

\begin{equation}
\Psi = \int\frac{1}{2} e^{2U} \mathcal{M}^{MN}\mathcal{Q}_{N}d\tau\, .
\end{equation}

}

\begin{eqnarray}
\label{eq:1dEqU}
e^{-W}\left[e^{W}U^{\prime}\right]^{\prime}+e^{2(U-W)}V_{\rm bh}+\frac{1}{2}e^{-2(U-W)}V &=& 0\, , \\
\label{eq:1dEqW}
e^{-W}\left[e^{W}\right]^{\prime\prime}+e^{-2(U-W)}V &=& 0\, , \\
\label{eq:1dEqphi}
e^{-W}\left[e^{W}\mathcal{G}_{ij}\phi^{j\, \prime}\right]^{\prime}-\frac{1}{2}\partial_i \mathcal{G}_{jk}\phi^{j\, \prime}\phi^{k\, \prime}+e^{2(U-W)}\partial_{i}V_{\rm bh}-\frac{1}{2}e^{-2(U-W)}\partial_{i}V  &=& 0\label{eq:scalars}\, ,
\end{eqnarray}

\noindent
to which we have to add the Hamiltonian constraint (\ref{eq:hamiltonian}). The kinetic term for the scalars, as well as the scalar potential $V(\phi)$ and the black hole potential $V_{\rm bh}(\phi,\mathcal{Q})$, can be solely expressed in terms of $U$ and  $W,$ i.e.,

\begin{eqnarray}
\label{eq:Visolated}
V &=&-e^{2 U-2 W} \left[W^{\prime\prime}+W^{\prime\, 2}\right] \, ,\nonumber\\
V_{\rm bh}(\phi,\mathcal{Q})&=&-\frac{1}{2} e^{2 W-2 U} \left[2 U^{\prime\prime}+2 U^{\prime} W^{\prime}-W^{\prime\prime}-W^{\prime\, 2}\right]\, ,\\
 \mathcal{G}_{ij}\phi^{i\, \prime}\phi^{j\, \prime}&=&-2 \left[-U^{\prime\prime}-U^{\prime} W^{\prime}+U^{\prime\, 2}+W^{\prime\prime}\right] \nonumber\, .
\end{eqnarray}

\noindent
Eqs. (\ref{eq:Visolated}) are useful in order to obtain, given a particular metric, the behavior of different quantities, like $V(\phi)$ and $V_{\rm bh}(\phi,\mathcal{Q})$, or $\phi^{i}$ for models with small enough number of scalars, in terms of the coordinate $r$. Of course, only metrics compatible with the equations of motion will yield consistent results.


\subsection{Constant scalars: generalities}


For constant scalars $\phi^{i}$, the potential $V(\phi)$ and the black hole potential $V_{\rm bh}(\phi,\mathcal{Q})$ become constant quantities, the former playing the role of a cosmological constant and the latter of a \emph{generalized} squared charge, magnetic and electric. In the case of constant scalars, Eq. (\ref{eq:1dEqphi})  is not identically satisfied, but it becomes the following constraint

\begin{eqnarray}
\label{eq:scalarconstraint}
e^{4(U-W)}\partial_{i}V_{\rm bh} = \frac{1}{2}\partial_{i}V \, .
\end{eqnarray}

\noindent
We have two different options in order to fulfil Eq. (\ref{eq:scalarconstraint}).


\paragraph{Constant scalars as \emph{double} critical points:~~ $\partial_{i}V_{\rm bh}=0,~~\partial_{i}V = 0$.}


Of course, the system of equations given by

\begin{eqnarray}
\label{eq:doublecritical}
\partial_{i}V_{\rm bh}=0,~~\partial_{i}V = 0\, ,
\end{eqnarray}

\noindent
is overdetermined. However, let's assume that a consistent solution to (\ref{eq:doublecritical}) exists and is given by

\begin{eqnarray}
\label{eq:criticalphi}
\phi^{i} = \phi^{i}_{c}\left(\mathcal{Q},\phi_{\infty}\right)\, ,
\end{eqnarray}

\noindent
\emph{i.e.}, the values of the scalars are fixed in terms of the electric and magnetic charges, and we have included a dependence on $\phi_{\infty}$ to formally consider the existence of flat directions. We will see later on that, in fact, Eq.~(\ref{eq:doublecritical}) happens in $\mathcal{N}=2,~d=4$ Supergravity. The equations of motion reduce to

\begin{eqnarray}
\label{eq:1dEqUcted}
e^{-W}\left[e^{W}U^{\prime}\right]^{\prime}+e^{2(U-W)}V_{\rm bh}+\frac{1}{2}e^{-2(U-W)}V &=& 0,\\
\label{eq:1dEqWcted}
e^{-W}\left[e^{W}\right]^{\prime\prime}+e^{-2(U-W)}V &=& 0\, ,
\end{eqnarray}

\noindent
together with the hamiltonian constraint

\begin{equation}
\label{eq:hamiltoniancted}
 2 U^{\prime\, 2}-2 W^{\prime\, 2}
+2e^{2(U-W)}V_{\rm bh}
-e^{-2(U-W)}V
=
0\, .
\end{equation}

\noindent


\paragraph{Metric functions identified:~~ $e^{U}=\beta e^{W},~~\beta\in\mathbb{R}^{+}$~~ and ~~ $2\beta^4\partial_{i}V_{\rm bh}=\partial_{i}V(\phi) $ .}


In this case, the equations of motion imply

\begin{equation}
\label{eq:constraintVV}
2\beta^4 \partial_{i}V_{\rm bh} = \partial_{i}V \, ,~~~ 2\beta^4 V_{\rm bh} = V \, .
\end{equation}

\noindent
Assuming Eqs.~(\ref{eq:constraintVV}), there is a unique solution, which is precisely $aDS_{2}\times\mathbb{R}^2$. Eqs. (\ref{eq:constraintVV}) can be understood as necessary and sufficient conditions for a gravity theory coupled to scalars and vector fields, up to two derivatives, to contain an $aDS_{2}\times \mathbb{R}^2$ solution. Therefore, given a particular theory of such kind, with a specific potential $V(\phi)$ and black hole potential $V_{\rm bh}(\phi)$, one only has to impose Eqs. (\ref{eq:constraintVV}) in order to check the existence of an $aDS_{2}\times \mathbb{R}^2$ solution.
\noindent
The parameter $\beta$ can be always found to be

\begin{equation}
\beta^4 = \frac{V}{2V_{\rm bh}} \, ,
\end{equation}

\noindent
and we are left with

\begin{equation}
\label{adsconstant}
\frac{1}{2} \partial_{i}\log V_{\rm bh} = \partial_{i}\log V \, .
\end{equation}

\noindent
Eq. (\ref{adsconstant}) is a system of $n_v$ equations for at least $n_v$ variables (the $n_v$ constant scalars), and hence in general it will be compatible and the theory will contain an $aDS_{2}\times \mathbb{R}^2$ solution. Only in pathological cases the system (\ref{adsconstant}) will be incompatible and the theory will fail to contain an $aDS_{2}\times \mathbb{R}^2$ solution.


\section{The Einstein-Maxwell-Dilaton model}
\label{sec:EMD}


Before we discuss the possible embeddings of Eq. (\ref{eq:action}) in gauged Supergravity and String Theory, let's consider the Einstein-Maxwell-Dilaton (E.M.D.) system, whose action is characterized by the following choices, to be made in Eq.~(\ref{eq:action})

\begin{eqnarray}
\label{eq:choicesEMD}
F^{\Lambda\, \mu\nu} = F^{ \mu\nu}, ~~~~I_{\Lambda\Sigma} = I =\frac{Z(\phi)}{2} < 0,~~~~R_{\Lambda\Sigma}= R =0,~~~~\phi^i = \phi,~~~~G_{ij}=\frac{1}{2}.\,
\end{eqnarray}

\noindent
Hence, the E.M.D. action reads
\begin{equation}
\label{eq:EMD}
S_{EMD}
 =
\int d^{4}x \sqrt{|g|}
\left\{
R
+\frac{1}{2} \partial_{\mu}\phi \partial^{\mu}\phi
+Z(\phi) F^{2}-V(\phi)\right\}\, ,
\end{equation}

\noindent
\emph{i.e.}, we consider a single vector field and a single scalar field. Moreover, the coupling given by $R$ is taken to be zero, which greatly simplifies the black hole potential $V_{\rm bh}(\phi,\mathcal{Q})$, which is therefore given by

\begin{eqnarray}
\label{eq:EMDVbh}
V_{\rm bh}(\phi,\mathcal{Q}) = \frac{1}{4}\left[Z(\phi) p^2 + Z(\phi)^{-1} q^2\right]\, ,
\end{eqnarray}

\noindent
where $q$ and $p$ are the electric and magnetic charges, respectively. The equations of motion take the form

\begin{eqnarray}
\label{eq:1dEqUemd}
e^{-W}\left[e^{W}U^{\prime}\right]^{\prime}+e^{2(U-W)}\frac{1}{4}\left[Z p^2 + Z^{-1} q^2\right]+\frac{1}{2}e^{-2(U-W)}V &=& 0\, ,\\
\label{eq:1dEqWemd}
e^{-W}\left[e^{W}\right]^{\prime\prime}+e^{-2(U-W)}V &=& 0\, ,\\
\label{eq:1dEqphiemd}
e^{-W}\left[e^{W}\phi^{\prime}\right]^{\prime}+e^{2(U-W)}\frac{\partial_{\phi}Z}{2}\left[ p^2 - \frac{q^2}{Z^{2}} \right]-e^{-2(U-W)}\partial_{\phi}V  &=& 0\, ,
\end{eqnarray}

\noindent
and the hamiltonian constraint reads

\begin{equation}
\label{eq:hamiltonianemd}
 2 U^{\prime\, 2}-2 W^{\prime\, 2}
+ \frac{1}{2}\phi^{\prime\, 2}
+\frac{e^{2(U-W)}}{2} \left[Z p^2 + Z^{-1} q^2\right]
-e^{-2(U-W)}V
=
0\, .
\end{equation}

\noindent
For non-constant scalars, Eq. (\ref{eq:1dEqphiemd}) is automatically satisfied if

\begin{eqnarray}
V&=&-e^{2(U-W)}\left[{W^{\prime}}^2+W^{\prime\prime} \right],\\
{\phi^{\prime}}^2&=&4\left[-{U^{\prime}}^2+U^{\prime}W^{\prime}+U^{\prime\prime}-W^{\prime\prime} \right] ,
\end{eqnarray}
\noindent
and $Z$ is such that
\begin{eqnarray}
\label{Zpq}
Z&=& \frac{1}{p^2}\left[ \Upsilon \pm \sqrt{\Upsilon^2-p^2q^2} \right], ~ \text{if}~p,~q\neq 0,\\ \label{Zp}
Z&=& \frac{2\Upsilon}{p^2}~ \text{if}~q=0,~p\neq 0\\ \label{Zq}
Z&=& \frac{q^2}{2\Upsilon}~ \text{if}~p=0,
\end{eqnarray}

\noindent
where

\begin{equation}
\displaystyle \Upsilon=2V_{\rm bh}=e^{2(W-U)}\left[-2U^{\prime}W^{\prime}+{W^{\prime}}^2-2U^{\prime\prime}+W^{\prime\prime} \right].
\end{equation}

\noindent
 Theories with conventional and sensible matter have to satisfy the null-energy condition (NEC) $n_{\mu}n_{\nu}T^{\mu\nu}\geq 0$, where $n_{\mu}$ is an arbitrary null vector and $T^{\mu\nu}$ is the correspondent energy-momentum tensor. This condition translates, for the E.M.D. case, into the following constraints
\begin{equation}
\Upsilon\leq0,~ {\phi^{\prime}}^2\geq 0.
\end{equation}

\noindent
Hence, it is equivalent to the requirement of a semi-negative definite black hole potential, and a semi-positive definite kinetic term for the scalar field, compatible with the condition $Z\left(\phi\right)$.


\subsubsection*{Another coordinate system: $A-B-f$ coordinates.}

There is another system of coordinates which we will use along this paper, and that will be useful for different purposes. It is related to the $U-W$ system of coordinates by the following identifications:

\begin{eqnarray}
\label{eq:coordinatechange}
\left(\frac{dr}{d\tilde{r}}\right)^2 = f^{-1}(\tilde{r}),~~~~e^{2U}= e^{2(A(\tilde{r})+B(\tilde{r}))} f(\tilde{r}),~~~~e^{2W}=e^{4A(\tilde{r})+2B(\tilde{r})}f(\tilde{r})\, ,
\end{eqnarray}

\noindent
giving rise to the metric

\begin{eqnarray}
\label{eq:ABmetric}
ds^2_{f}=\ell^2 e^{2A(\tilde{r})} \left[ e^{2B(\tilde{r})} f(\tilde{r}) dt^2 - \frac{d\tilde{r}^2}{f(\tilde{r})}-\delta_{ij}dx^{i} dx^{j}\right]\, ,
\end{eqnarray}

\noindent
which has proven to be useful (see e.g. \cite{Iizuka:2011hg}, \cite{Dong:2012se}) in order to obtain solutions exhibiting \emph{hvLif} asymptotics when $f(\tilde{r})$ is a function of $\tilde{r}$ that obeys

\begin{equation}
f \left(\tilde{r}_{h}\right)=0,~~\tilde{r}_{h}\in \mathbb{R}^{+}~~~~~~\lim_{\tilde{r}\rightarrow \tilde{r}_0} f\left(\tilde{r}\right)=1\, .
\end{equation}

\noindent
The \emph{hvLif} limit is, thus, assumed to be at $\tilde{r}_0$, whereas the horizon is at $\tilde{r}_h$. The equations of motion (\ref{eq:1dEqUemd}), (\ref{eq:1dEqWemd}) and  (\ref{eq:1dEqphiemd}) can be rewritten accordingly as\footnote{From now on, we will use always the symbol "$r$" to denote the "radial" coordinate, independently of which coordinate system we use, which will be specified by other means.}

\begin{eqnarray}
\label{eq:1dEqUAB}
e^{-2A-B}\left[e^{2A+B}f \left[A^{\prime}+B^{\prime}+\frac{f^{\prime}}{2 f}\right]\right]^{\prime}+e^{-2A} \frac{1}{4}\left[Z p^2 + Z^{-1} q^2\right]+\frac{1}{2}e^{2A}V(\phi) &=& 0\\
\label{eq:1dEqWAB}
e^{-2A-B}\left[f^{1/2}\left[ e^{2A+B}f^{1/2}\right]^{\prime}\right]^{\prime}+e^{2A}V(\phi) &=& 0\\
\label{eq:1dEqphiAB}
e^{-2A-B}\left[e^{2A+B}f\phi^{\prime}\right]^{\prime}+e^{-2A} \frac{\partial_{\phi}Z}{2}\left[ p^2 -Z^{-2}q^2 \right]-e^{2A} \partial_{\phi}V(\phi)  &=& 0\, ,
\end{eqnarray}

\noindent
where $ ^{\prime} = \frac{d}{d\tilde{r}}$. The Hamiltonian constraint is given by

\begin{equation}
\label{eq:hamiltonianAB}
- 2f\left[3{A^{\prime}}^2+2A^{\prime}\left[B^{\prime}+\frac{f^{\prime}}{2f}\right]\right]
+ \frac{f}{2}\phi^{\prime\, 2}
+\frac{e^{-2A}}{2} \left[Z p^2 + Z^{-1} q^2\right]
-e^{2A}V(\phi)
=
0\, .\end{equation}
\noindent
Again, for non-constant dilaton this set of equations is equivalent to\footnote{Eqs. (\ref{Zpq}), (\ref{Zp}), (\ref{Zq}) hold.}

\begin{eqnarray}
\label{eqpVabf}
V&=& \frac{e^{-2A}}{2}\left[-3f^{\prime}[2A^{\prime}+B^{\prime}]-2f\left[2A^{\prime\prime}+\left[2A^{\prime}+B^{\prime} \right]^2+B^{\prime\prime}\right]-f^{\prime\prime} \right],\\\label{eqphiabf}
{\phi^{\prime}}^2&=&4\left[-A^{\prime\prime}+A^{\prime}B^{\prime}+{A^{\prime}}^2\right],\\\label{eqUPabf}
\Upsilon&=&-\frac{e^{2A}}{2}\left[f^{\prime}\left[2A^{\prime}+3B^{\prime} \right]+2f\left[2A^{\prime}B^{\prime}+B^{\prime\prime}+{B^{\prime}}^2\right]+f^{\prime\prime} \right].
\end{eqnarray}


\section{$\mathcal{N}=2$ Supergravity with F.I. terms}
\label{sec-sugramodel}


The action (\ref{eq:action}) has great generality and basically covers any possible theory of gravity coupled to abelian vector fields and scalars up to two derivatives. However, in order to embed our results in String Theory, it is convenient to focus on the bosonic sector of $\mathcal{N}=2,~d=4$ Supergravity, which is a particular case of (\ref{eq:action}). More precisely, we are going to consider gauged $\mathcal{N}=2,~d=4$ in the presence of $n_{v}$ abelian vector multiplets, where the gauge group is contained in the $R$-symmetry group of automorphisms of the supersymmetry algebra. Normally one refers to this theory as $\mathcal{N}=2,~d=4$ Supergravity with Fayet-Iliopoulos terms (from now on, $\mathcal{N}=2 ~\rm FI$ to abridge) \cite{Andrianopoli:1996cm}. The general lagrangian of $\mathcal{N}=2 ~\rm FI$ is given by

\begin{equation}
\label{eq:generalactionsugra}
\begin{array}{rcl}
 S & = & {\displaystyle\int} d^{4}x \sqrt{|g|}
\left\{R +2\mathcal{G}_{ij^{*}}\partial_{\mu}z^{i}
\partial^{\mu}z^{*\, j^{*}}
+2\Im{\rm m}\mathcal{N}_{\Lambda\Sigma}
F^{\Lambda\, \mu\nu}F^{\Sigma}{}_{\mu\nu}
 \right. \\
& & \\
& & \left.
\hspace{2cm}
-2\Re{\rm e}\mathcal{N}_{\Lambda\Sigma}
F^{\Lambda\, \mu\nu}{}^{\star}F^{\Sigma}{}_{\mu\nu}
-V_{\rm fi}\left(z,z^{*}\right)
\right\}\, .
\end{array}
\end{equation}

\noindent
The indices $i,j,\dotsc=1,\dotsc ,n_{v}$ run over the scalar fields and the indices $\Lambda,\Sigma,\dotsc=0,\dotsc , n_{v}$ over the 1-form
fields. The scalar potential generated by the F.I. terms reads

\begin{equation}
\label{eq:potentialFIgeneral}
V_{\rm fi}\left(z,z^{*}\right) =-3|\mathcal{Z}_{g}|^2+\mathcal{G}^{ij^{*}}\mathfrak{D}_i \mathcal{Z}_{g}\mathfrak{D}_{j^{*}} \mathcal{Z}^{*}_{g},~~~~\mathfrak{D}_i \mathcal{Z}_{g}=\partial_i \mathcal{Z}_{g}+\frac{1}{2}\partial_i \mathcal{K} \mathcal{Z}_{g}\, ,
\end{equation}

\noindent
where $\mathcal{K}$ is the K\"ahler potential, $\mathcal{Z}_{g}$ is given by\footnote{We assume the conventions of \cite{Meessen:2006tu}.}

\begin{equation}
\label{eq:tildeZ}
\mathcal{Z}_{g}\equiv \mathcal{Z}_{g}\left(z,z^{*}\right) = g_{M} \mathcal{V}^{M}=\mathcal{V}^{M}g^N\Omega_{MN}
=
-g^{\Lambda}\mathcal{M}_{\Lambda} +g_{\Lambda}\mathcal{L}^{\Lambda}\, ,
\end{equation}

\noindent
and the $g^M$ is a symplectic vector related to the embedding tensor $\theta_M$, that selects the combination of
vectors that gauges $U(1)\subset R$-symmetry group, as follows\footnote{Supergravity gaugings are originally electric, breaking therefore the symplectic covariance present in the ungauged case. The embedding tensor formalism allows to formally keep the theory simplectically covariant by introducing magnetic and electric gaugings. }

\begin{equation}
\label{eq:etensor}
g_M = g \theta_M\, ,
\end{equation}

\noindent
$g$ being the gauge coupling constant. The corresponding one-dimensional effective action and the hamiltonian constraint are given, respectively, by

\begin{equation}
\label{eq:effectiveactionsugra}
  \begin{array}{rcl}
S
=
{\displaystyle
\int dr\, e^{W}\,
\left\{
U^{\prime\, 2}-W^{\prime\, 2}
+ \mathcal{G}_{ij^{*}}z^{i\, \prime}z^{j^{*}\, \prime}
-e^{2(U-W)}V_{\rm bh}
+\frac{1}{2}e^{-2(U-W)}V_{\rm fi}
\right\}\, ,
}
\end{array}
\end{equation}

\begin{equation}
\label{eq:hamiltoniansugra}
 U^{\prime\, 2}-W^{\prime\, 2}
+ \mathcal{G}_{ij^{*}}z^{i\, \prime}z^{j^{*}\, \prime}
+e^{2(U-W)}V_{\rm bh}
-\frac{1}{2}e^{-2(U-W)}V_{\rm fi}
=
0\, .
\end{equation}

\noindent
The black-hole potential takes the simple form

\begin{equation}
\label{eq:sugraVbh}
-V_{\rm bh}(z,z^{*},\mathcal{Q}) = |\mathcal{Z}|^{2}
+\mathcal{G}^{ij^{*}}\mathfrak{D}_{i}\mathcal{Z}\mathfrak{D}_{j^{*}}\mathcal{Z}^{*}\, ,
\end{equation}

\noindent
where

\begin{equation}
\label{centralcharg}
\mathcal{Z}=\mathcal{Z}(z,z^{*},\mathcal{Q})  \equiv
\langle \mathcal{V}\mid\mathcal{Q} \rangle =
-\mathcal{V}^{M}\mathcal{Q}^{N}\Omega_{MN}
=
p^{\Lambda}\mathcal{M}_{\Lambda} -q_{\Lambda}\mathcal{L}^{\Lambda}\, ,
\end{equation}

\noindent
is the \textit{central charge} of the theory.


\paragraph{Constant scalars and supersymmetric attractors.}

In section (\ref{sec:effectiveaction}) we studied the case of constant scalars in the general theory (\ref{eq:action}). We found that, besides the solution $aDS_2\times\mathbb{R}^2$, there was another possible solution, if Eq. (\ref{eq:doublecritical}) holds. We will see now how this is always possible in $\mathcal{N}=2 ~\rm FI$. The general theory of the attractor mechanism in ungauged $d=4$ Supergravity proves that, for extremal black holes, the value of the scalars at the horizon is fixed in terms of the charges $\mathcal{Q}^M$, and given by the so called \emph{critical points} or \emph{attractors}, \emph{i.e.}, solutions to the system

\begin{equation}
\label{eq:crticialun}
\partial_{i}V_{\rm bh}\left(\mathcal{Q},\phi\right)_{\left. \right|_{\phi_c}}=0\, .
\end{equation}

\noindent
There might be some residual dependence in the value at infinity if the potential has flat directions. If the scalars are constant, they have to be given again by (\ref{eq:crticialun}) in the extremal as well as in the non-extremal case. It can be proven that there is always a class of attractors, called supersymmetric, which obey

\begin{equation}
\label{eq:crticialunsusy}
\partial_{i}\left|\mathcal{Z}\right|_{\left. \right|_{\phi_c}}=0\, ,~~~{\rm and} ~~~\mathfrak{D}_i\mathcal{Z}_{\left. \right|_{\phi_c}} = 0\, ,
\end{equation}

\noindent
and therefore, given the definitions (\ref{eq:potentialFIgeneral}) and (\ref{eq:sugraVbh}), they also obey (\ref{eq:doublecritical}) if $\mathcal{Q}^M\sim g^M$. Hence, setting the scalars to constant values given by the supersymmetric attractor points of the black hole potential is always a consistent truncation, provided that $g^M$ is identified with $\mathcal{Q}^M$, which besides fixes the value of the black hole potential and the scalar potential exclusively in terms of the charges.


\subsection{The $t^3$-model}
\label{sec-STmodel}


In this section we consider a particular $\mathcal{N}=2 ~\rm FI$ model which can be embedded in String Theory. In particular we start from Type-IIB String Theory compactified on a Sasaki-Einstein manifold to five dimensions. This theory can be consistently truncated as to yield pure $\mathcal{N}=1,~d=5$ Supergravity with Fayet-Iliopoulos terms, which, due to the absence of scalars, introduce a cosmological constant. Further compactification on $S^1$ gives us the desired four dimensional theory, which is defined by \cite{Cassani:2010uw,Cassani:2011sv,Liu:2010sa,Gauntlett:2010vu}

\begin{equation}
\label{eq:embedding}
n_{v}=1,~~~ F\left( \mathcal{X}\right) = -\frac{\left(\mathcal{X}^1\right)^3}{\mathcal{X}^0},~~~g^0 = g^1 = g_0 = 0\Rightarrow V_{\rm fi}\left(t,t^{*}\right)=\frac{-\beta^2}{\Im{\rm m}t}\, ,
\end{equation}

\noindent
where $\beta^2=g^2_1/3$, and we have defined the inhomogeneous coordinate on the Special K\"ahler manifold ${\rm SU}(1,1)/{\rm U}(1)$, by

\begin{equation}
t=\frac{\mathcal{X}^1}{\mathcal{X}^0}\,.
\end{equation}

\noindent
This theory is known as the $t^3$-model, and although the String Theory embedding requires the gauging specified in Eq. (\ref{eq:embedding}), we are going to study it in full generality, particularizing only at the end.

\noindent
The canonically normalized symplectic section $\mathcal{V}$ is, in a certain
gauge,

\begin{equation}
\mathcal{V}=e^{\mathcal{K}/2}
\left(
  \begin{array}{c}
1 \\ t \\ t^3 \\ -3t^2 \\
  \end{array}
\right)\, ,
\end{equation}

\noindent
where the K\"ahler potential is

\begin{equation}
\mathcal{K} \,=\, -\log \left[\,i \left(t-t^{*}\right)^3\right] \,.
\end{equation}

\noindent
As a consequence, the K\"ahler metric reads

\begin{equation}
\mathcal{G}_{tt^{*}} \,=\, \frac{3}{4}\,\frac{1}{(\Im{\rm m}t)^2} \,,
\end{equation}

\noindent
and the central charge

\begin{equation}
\mathcal{Z}=\frac{p^0 t^3-3t^2p^1-q_0-q_1 t}{2\sqrt{2\Im mt^3}}\,.
\end{equation}

\noindent
The period matrix $\mathcal N_{IJ}$ is, in turn, given by

\begin{equation}
{\rm Re}\,\mathcal N_{IJ} \,=\, \left(\begin{array}{cc}
-2\Re^3  &   3 \Re^2\\ [2mm]
3 \Re^2  &  -6 \Re
\end{array}\right)\;,\qquad {\rm Im}\,\mathcal N_{IJ} \,=\, \left(\begin{array}{cc}
-(\Im^3 + 3 \Re^2 \Im)  & 3\Re \Im  \\ [2mm]
 3\Re \Im  &  -3\Im
\end{array}\right)\,,
\end{equation}

\noindent
where we use the notation: $\Re \equiv \Re{\rm e} t,~~\Im \equiv \Im{\rm m} t$. The general expressions of $V_{\rm bh}$ and $V_{\rm fi}$, which can be obtained using Eqs. (\ref{eq:potentialFIgeneral}), (\ref{eq:tildeZ}), (\ref{eq:sugraVbh}) and (\ref{centralcharg}) read

\begin{align}
V_{\rm bh}=& -\frac{1}{6 \Im^3}\left[3 \Im^6 {p^0}^2+9 \Im^4 \left[p^1-p^0 \Re\right]^2+\Im^2 \left[q_1+6p^1\Re-3p^0\Re^2 \right]^2 \right.\\ & \notag  \left. +3\left[q_0+\Re \left[q_1+3p^1\Re -p^0 \Re^2 \right]\right]^2   \right],\\
V_{\rm fi}=& -\frac{1}{3 \Im} \left[g_1^2+3g_1\left[{g^1}\Re +g^0 \left[\Im^2+\Re^2 \right] \right]+9 \left[g_0 \left[-g^1+g^0\Re \right]+{g^1}^2 \left[\Im^2+\Re^2 \right] \right] \right].
\end{align}

\noindent
Let's consider the truncation $\Re{\rm e}t=0$. In order to satisfy all the original equations of motion (those with  $\Re{\rm e}t$ arbitrary) in such a case, we must impose the additional constraints

\begin{equation}
\partial_{\Re}V_{\rm bh}(\Re=0)=\partial_{\Re}V_{\rm fi}(\Re=0)=0.
\end{equation}

\noindent
These conditions explicitly read

\begin{equation}
3\Im p^0 p^1-2\frac{p^1 q_1}{\Im}-\frac{q_0 q_1}{\Im^3}=0,
\end{equation}
\begin{equation}
3g_0g^0+g_1 g^1=0,
\end{equation}

\noindent
and are satisfied (without loss of generality in the functional form of the potentials) if we make

\begin{equation}
p^1=q_1=0;~~g_0=g^1=0 \vee g^0=g_1=0.
\end{equation}

\noindent
Thus, setting $\Re{\rm e}t$ to zero in a consistent manner notably simplifies the expressions for the potentials
\begin{equation}
V_{\rm bh}= -\frac{1}{2}\left[\frac{q_0^2}{\Im^3}+{p^0}^2 \Im^3 \right],
\end{equation}
\begin{equation}
V^I_{\rm fi}= -\left[\frac{g_1^2}{3\Im}+g_1 g^0 \Im \right],~V^{II}_{\rm fi}=- \left[-\frac{3g_0g^1}{\Im}+3{g^1}^2 \Im \right]
\end{equation}

\noindent
The action is, making the redefinition $t\equiv \Re + i e^{-\frac{\phi}{\sqrt{3}}}$, given by

\begin{equation}
\label{eq:t3Rt0}
S^I_{\Re = 0}
 =
\int d^{4}x \sqrt{|g|}
\left\{
R
+\frac{1}{2} \partial_{\mu}\phi \partial^{\mu}\phi
-2 e^{-\sqrt{3}\phi} \left(F^{0}\right)^{2}
+\frac{g_1^2}{3}e^{\frac{\phi}{\sqrt{3}}}+g_1g^0 e^{-\frac{\phi}{\sqrt{3}}}
\right\}\, ,
\end{equation}

\begin{equation}
\label{eq:t3Rt02}
S^{II}_{\Re = 0}
 =
\int d^{4}x \sqrt{|g|}
\left\{
R
+\frac{1}{2} \partial_{\mu}\phi \partial^{\mu}\phi
-2 e^{-\sqrt{3}\phi} \left(F^{0}\right)^{2}
-3g_0g^1e^{\frac{\phi}{\sqrt{3}}}+3{g^1}^2 e^{-\frac{\phi}{\sqrt{3}}}
\right\}\, ,
\end{equation}

\noindent
where we have already set $A^{1}_{\mu}$ to zero, in order to make the truncation consistent with the corresponding equation of motion.

\paragraph{Embedding the $t^3$-model system in the E.M.D.} As it can be trivally verified, we have just obtained the action (\ref{eq:EMD}) with

\begin{equation}
Z(\phi)=-2e^{-\sqrt{3}\phi},~~q^2=4q_0^2,~~p^2={p^0}^2,\,
\end{equation}

\noindent
and the scalar potential of the E.M.D. system (Eq. (\ref{eq:Vc_2})) given by
\begin{equation}
V(\phi)=c_1 e^{-\frac{\phi}{\sqrt{3}}}+c_2 e^{+\frac{\phi}{\sqrt{3}}};~~c^I_1=- g_1 g^0,~~c^{II}_1=-3{g^1}^2,~~c^I_2=- \frac{g_1^2}{3},~~c^{II}_2=3g_0g^1.\,
\end{equation}

\noindent
Hence, we find that our axion-free $t^3$-system with those particular choices of $Z$ and $V$ gets embedded in the E.M.D. model and, for $g^0=g^1=g_0=0$, also in String Theory in the way explained at the beginning of this section. In such a case, Eq. (\ref{eq:t3Rt0}) clearly becomes

\begin{equation}
\label{eq:t3Rt0ST}
S_{ST}
 =
\int d^{4}x \sqrt{|g|}
\left\{
R
+\frac{1}{2} \partial_{\mu}\phi \partial^{\mu}\phi
-2 e^{-\sqrt{3}\phi} \left(F^{0}\right)^{2}
+\frac{g_1^2}{3}e^{\frac{\phi}{\sqrt{3}}}
\right\}\, .
\end{equation}


\section{\emph{hvLif} solutions}


In this section we are going to construct (purely and asymptotically) \emph{hvLif} solutions to Eq. (\ref{eq:action}). After establishing some results on the properties of the solutions corresponding to the pure \emph{hvLif} case in the general set-up of Eq. (\ref{eq:Visolated}), we focus on the E.M.D. system, obtaining the \emph{hvLif} solutions allowed by the embedding of our \emph{axion-free} Supergravity model in this system. Then, we provide a recipe to construct asymptotically \emph{hvLif} solutions to these theories in the presence of constant and non-constant dilaton fields, recovering and extending some of the results already present in the literature.


\subsection{Purely \emph{hvLif} solutions: general remarks}
\label{sec:purelyhvLf}


The \emph{hvLif} metric in four dimensions, given by

\begin{eqnarray}
\label{eq:purelyLif}
ds^2 = \ell^2 r^{\theta-2}\left(r^{-2(z-1)}dt^2-dr^2-\delta_{ij}dx^i dx^j\right)\, ,
\end{eqnarray}

\noindent
is recovered in our set-up for specific values of $U(r)$ and $W(r)$, namely

\begin{equation}
\label{eq:hvLifUR}
e^{2U(r)} = \ell^2 r^{\theta-2z},~~~ e^{2W(r)} = \ell^4 r^{2(\theta-z-1)}\, .
\end{equation}

\noindent
For purely \emph{hvLif} solutions, the equations of motion can be further simplified by direct substitution of (\ref{eq:hvLifUR})

\begin{eqnarray}
\label{eq:1dEqUHv}
(\theta-2z)(\theta-z-2)+2 r^{4-\theta}\ell^{-2} V_{\rm bh}+r^{\theta}\ell^{2} V &=& 0\\
\label{eq:1dEqWHv}
(\theta-z-1)(\theta-z-2)+r^{\theta}\ell^{2}V &=& 0\\
\label{eq:1dEqphiHv}
r^{-2(\theta-z-1)}\left(r^{2(\theta-z-1)}\mathcal{G}_{ij}\phi^{j\, \prime}\right)^{\prime}-\frac{1}{2}\partial_i \mathcal{G}_{jk}\phi^{j\, \prime}\phi^{k\, \prime}+r^{2-\theta}\ell^{-2}\partial_{i}V_{\rm bh} - \frac{1}{2}r^{\theta-2}\ell^{2}\partial_{i}V  &=& 0\, .
\end{eqnarray}

\noindent
The Hamiltonian constraint reads

\begin{equation}
\label{eq:hamiltonianHv}
(2-\theta)(3 \theta - 4 z-2)
+ 2r^2\mathcal{G}_{ij}\phi^{i\, \prime}\phi^{j\, \prime}
+4r^{4-\theta}\ell^{-2}V_{\rm bh}
-2r^{\theta}\ell^{2}V
=
0\, .
\end{equation}

\noindent
Eqs. (\ref{eq:Visolated}) can be also adapted to the purely \emph{hvLif} case. We find

\begin{eqnarray}
\label{eq:VVbehave}
V = -\ell^{-2} \mathcal{X}_{(\theta,z)}r^{-\theta} ,~~~~
V_{\rm bh}(\phi,\mathcal{Q}) = \frac{1}{2} \ell^2  \mathcal{Y}_{(\theta,z)} r^{\theta-4},~~~~
 \mathcal{G}_{ij}\dot{\phi}^{i}\dot{\phi}^{j} = \frac{1}{2}\mathcal{W}_{(\theta,z)}r^{-2}\, ,
\end{eqnarray}

\noindent
where

\begin{eqnarray}
\mathcal{X}_{(\theta,z)}&=&(\theta-z-2) (\theta-z-1), \\
\mathcal{Y}_{(\theta,z)}&=&(\theta-z-2)(z-1),\\
\mathcal{W}_{(\theta,z)}&=&(\theta-2) (\theta-2 z+2)\, .
\end{eqnarray}

\noindent
Eqs. (\ref{eq:1dEqUHv})-(\ref{eq:hamiltonianHv}) are the general equations of motion that need to be solved in order to find a \emph{hvLif} solution to any theory that belongs to the class defined by Eq. (\ref{eq:action}). Likewise, Eq. (\ref{eq:VVbehave}) provides the behaviour of the black hole potential and the scalar potential, in terms of the variable $r$, for any \emph{hvLif} solution consistent with the equations of motion. $\mathcal{G}_{ij}$ is positive-definite, therefore
\begin{equation}
\mathcal{G}_{ij} \phi^{i\, \prime}\phi^{j\, \prime} \geq 0 \Leftrightarrow \mathcal{W}_{(\theta,z)}\geq 0,\qquad \mathcal{G}_{ij}\phi^{i\, \prime}\phi^{j\, \prime}=0~ \Leftrightarrow~\phi^{i\, \prime}=0~\forall i\, ,
\end{equation}

\noindent
and hence we can establish the following result: all the scalar fields of any purely \emph{hvLif} solution of any theory describable by Eq. (\ref{eq:action}) are constant \emph{iff} $\theta=2$, or $z=1+\theta/2$. In addition, $V_{\rm bh}$ is, in our conventions, a negative definite function, hence $V_{\rm bh}\leq 0\Leftrightarrow \mathcal{Y}_{(\theta,z)}\leq 0$. These two conditions on the sign of $\mathcal{W}_{(\theta,z)}$ and $ \mathcal{Y}_{(\theta,z)}$ are equivalent to imposing the null-energy condition (NEC) to our purely \emph{hvLif} solutions, as we commented before, and define a region of acceptable solutions in the $(\theta,z)$-plane, as we shall see.\\

\noindent
It is possible to stablish some general results for the \emph{hvLif} solutions of any theory describable by Eq. (\ref{eq:action}) attending to the vanishing of $V$, $V_{\rm bh}$ and/or $\mathcal{G}_{ij} \dot{\phi}^i\dot{\phi}^j$. Let's proceed.

\begin{enumerate}

\item $ {\theta=2}$\\
In this situation $\mathcal{G}_{ij} \dot{\phi}^i\dot{\phi}^j=0$, and
\begin{eqnarray}
V&=& -\ell^{-2} z(z-1) r^{-2},\\
V_{\rm bh}&=& -\frac{1}{2}\ell^2 z(z-1)r^{-2}.
\end{eqnarray}

\noindent
The NEC imposes ${z\in(-\infty,0]\cup[1,+\infty)}$, and we have the two special cases: ${\theta=2,~z=0}$ (which corresponds to Rindler spacetime) and ${\theta=2,~z=1}$ (which is Minkowski space-time) for which $V=V_{\rm bh}=0$ as well.

\item ${z=1+\frac{\theta}{2}}$\\
We have again $\mathcal{G}_{ij} \dot{\phi}^i\dot{\phi}^j=0$, and

\begin{eqnarray}
V&=& -\ell^{-2} \left(\frac{\theta}{2}-3\right)\left(\frac{\theta}{2}-2\right) r^{-\theta},\\
V_{\rm bh}&=& -\frac{1}{2}\ell^2 \left(\frac{\theta}{2}-3\right)\frac{\theta}{2}r^{\theta-4}.
\end{eqnarray}

\noindent
The NEC translates into ${\theta\in[0,6]}$, and we have three more special cases: the Ricci flat one: ${\theta=6,~z=4}$ corresponding to $V=V_{\rm bh}=0$ (this is a particular case of the general formalism developed in \cite{Bueno:2012sd} for ungauged $\mathcal{N}=2,~d=4$ Supergravity); ${\theta=4,~z=3}$, which corresponds to $V=0$, $V_{\rm bh}=-\ell^2$ (also in agreement with the results of \cite{Bueno:2012sd}); and ${\theta=0,~z=1}$, which is nothing but the $aDS_{4}$ space-time in a conformally flat representation, and the only solution with vanishing black hole potential, and constant (non-zero) scalar potential compatible with the equations: $V_{\rm bh}=0$, $V\equiv\Lambda=-\ell^{-2} 6$.

\item $ {z=1,~\theta\neq 2,~z\neq 1+\frac{\theta}{2}}$\\ We have $V_{\rm bh}=0$, whereas

\begin{eqnarray}
V&=& -\ell^{-2} \left(\theta-3\right)\left(\theta-2\right) r^{-\theta},\\
\mathcal{G}_{ij}\dot{\phi}^i\dot{\phi}^j&=&\frac{1}{2}(\theta-2)\theta ~ r^{-2}.
\end{eqnarray}

\noindent
The NEC becomes now ${\theta\in (-\infty,0]\cup[2,\infty)}$, and we have the limit case ${\theta=3,~z=1}$ which will be a particular case of the family considered in the next paragraph.

\item $ {z=\theta-2,~\theta\neq 2,~z\neq 1+\frac{\theta}{2}}$\\ This situation imposes $V=V_{\rm bh}=0$, whereas

\begin{eqnarray}
\mathcal{G}_{ij}\dot{\phi}^i\dot{\phi}^j&=&\frac{1}{2}(\theta-2)(6-\theta) ~ r^{-2}.
\end{eqnarray}

\noindent
The NEC reads ${\theta\in [2,6]}$. These will be solutions of the Einstein-Dilaton system for $\mathcal{G}_{ij}=\frac{1}{2}\delta_{ij},~i=1$, and

\begin{equation}
\phi=\phi_0+\sqrt{(\theta-2)(6-\theta)}\log r.
\end{equation}

\item $ {z=\theta-1,~\theta\neq 2,~z\neq 1+\frac{\theta}{2}}$\\ We have now $V=0$, while

\begin{eqnarray}
V_{\rm bh}&=& -\frac{1}{2}\ell^2 \left(\theta-2\right)\frac{\theta}{2}r^{\theta-4}.\\
\mathcal{G}_{ij}\dot{\phi}^i\dot{\phi}^j&=&\frac{1}{2}(\theta-2)(4-\theta) ~ r^{-2},
\end{eqnarray}

\noindent
and the NEC becomes ${\theta\in [2,4]}$.\\

\noindent

\end{enumerate}

Another particularly interesting case corresponds to the Einstein-Maxwell system with a cosmological constant: $\mathcal{G}_{ij}\dot{\phi}^i\dot{\phi}^j=0$, $V\equiv\Lambda$. However, this could only be realized for $\theta=0,~z=1$, which imposes the vanishing of $V_{\rm bh}$. Hence, there is no purely \emph{hvLif} solution (for non-vanishing vector fields) for such model.

\begin{figure}[ht]
\centerline{\includegraphics[scale=0.4]{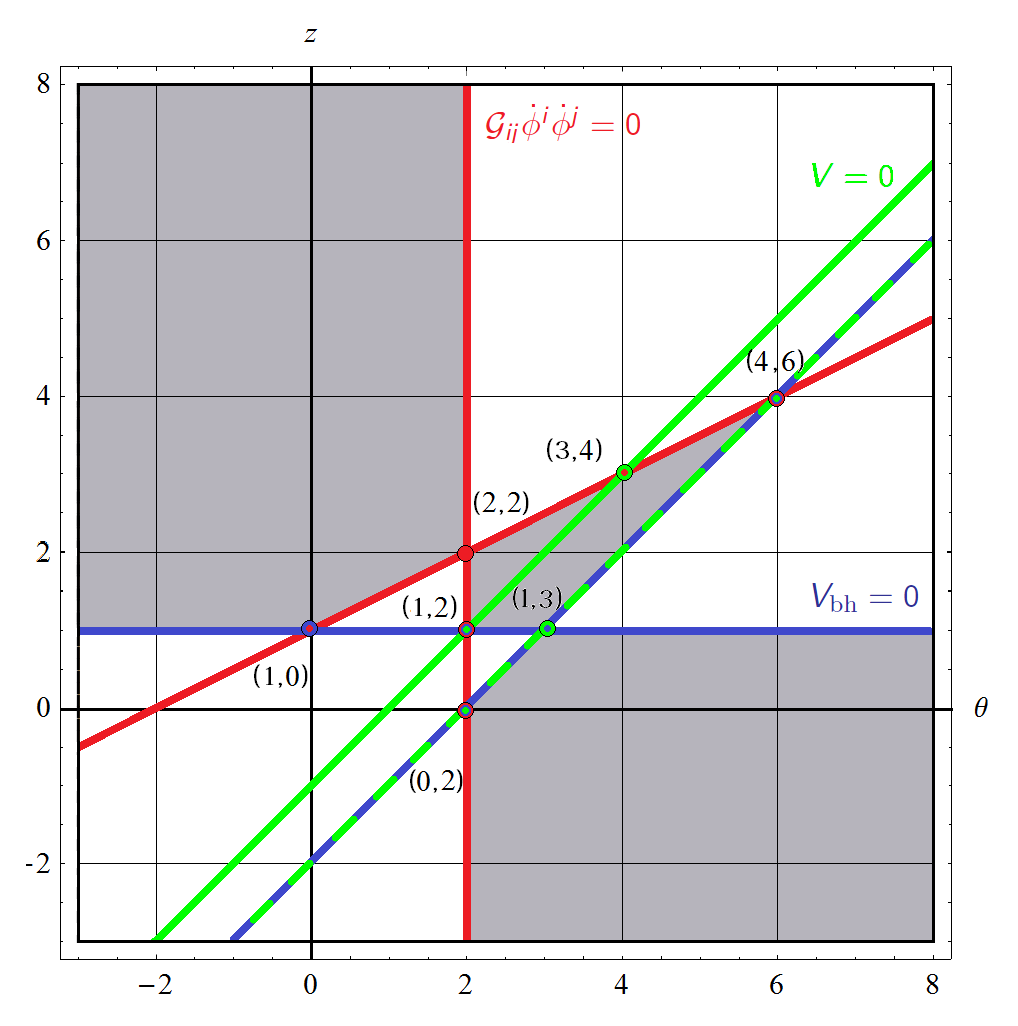}}
\caption{\textsl{Purely hvLif ($\theta,z$) plane. Red lines correspond to $\mathcal{G}_{ij}\dot{\phi}^i\dot{\phi}^j=0$, the blue ones to $V_{\rm bh}=0$, and those in green to $V=0$. The shaded regions represent solutions which satisfy the NEC.}}
\end{figure}


\subsubsection{Purely \emph{hvLif} in the E.M.D.}
If we particularize now to the E.M.D. system, we find
\begin{eqnarray}
V &=& -\ell^{-2} \mathcal{X}_{(\theta,z)}r^{-\theta} ,\\
\Upsilon&=&2V_{\rm bh}=\ell^2 \mathcal{Y}_{(z,\theta)} r^{\theta-4},\\ \label{diladila}
\phi&=&\phi_0+ \sqrt{\mathcal{W}_{(z,\theta)}} \log (r)~ \Rightarrow~ r=e^{\frac{\phi}{\sqrt{\mathcal{Z}}}}  \, .
\end{eqnarray}
\noindent
Therefore, $V$ and $V_{\rm bh}$ written as functions of $\phi$, must take the form

\begin{eqnarray}\label{Vemd}
V(\phi)&=& -\ell^2 \mathcal{X} e^{-\frac{\theta \phi}{\sqrt{\mathcal{Z}}}},\\
V_{\rm bh}(\phi)&=& \frac{1}{2}\ell^2 \mathcal{Y} e^{\frac{(\theta-4)\phi}{\sqrt{\mathcal{Z}}}}.
\end{eqnarray}
This means, on the one hand, that any E.M.D. theory susceptible of containing \emph{hvLif} solutions has a scalar potential which depends on $\phi$ through one single exponential (becoming a constant when $\theta=0$, $\theta=2$ or $z=1+\theta/2$ ($\phi=\phi_0$ in the last two cases)) \cite{Bhattacharya:2012zu}. On the other hand, the gauge coupling function is constant for $\theta=4$, and again if $\phi=\phi_0$.\\


\subsubsection*{$t^3$-model}

Let's see now what the situation is for the truncation of the $t^3$-model considered in the previous section. In this case, $V^{I,II}=c_1 e^{-\phi/\sqrt{3}}+c_2 e^{\phi/\sqrt{3}}$ with $c_2=0~\Rightarrow~c_1=0$ in the case I, and $c_1=0~\Rightarrow~c_2=0$ in the case II. Since we can only keep one of the exponentials (in order to match $V$ with Eq. (\ref{Vemd})), the only possibility is setting $g^0=0$ ($c_1=0$) in the case I (which leaves us with the String Theory embedded model), and $g_0$ ($c_2=0$) in the case II. In both situations, $Z(\phi)=-2~e^{-\sqrt{3}\phi}$. In I there exists one only solution, which is magnetic, and corresponds to ${\theta=-2,~z=3/2}$, $g_1^2=297/(4 \ell^2)$ and $p^2=11\ell^2/4$. On the other hand, case II admits one only solution (magnetic as well) for ${\theta=1,~z=3}$, ${g^1}^2=4/ \ell^2$, and $p^2=8\ell^2$. Both solutions satisfy the NEC, as it was desirable, and have a running dilaton given by Eq. (\ref{diladila}) with $\mathcal{Z}=12$ and $\mathcal{Z}=3$ respectively.

\subsection{Asymptotically \emph{hvLif} in the E.M.D.}
\subsubsection{Non-constant scalar field}

In order to construct new solutions with \emph{hvLif} asymptotics, we switch now to $A-B-f$ variables. The required form for $A$ and $B$ is
\begin{equation}
\label{ansatz}
e^{2A}=r^{\theta-2},~~e^{2B}=r^{-2(z-1)}.
\end{equation}
With this election, Eq. (\ref{eqphiabf}) can be directly integrated, yielding
\begin{equation}
\label{eq:phi0}
\phi=\phi_0+\sqrt{(\theta-2)(\theta+2-2z)}\log(r).
\end{equation}
$\Upsilon$ and $V$, in turn, become\footnote{Recall that $Z$ is given in term of $\Upsilon$ in Eqs. (\ref{Zpq}), (\ref{Zp}) and (\ref{Zq}) depending on the case}
\begin{eqnarray}
\label{eqV1}V&=&\frac{1}{2}r^{-\theta}\left[\left[1-\theta+z\right]\left[2\left[\theta-2-z \right]f+3r f^{\prime} \right]-r^2 f^{\prime\prime} \right], \\
\label{eqZ1}\Upsilon&=&r^{\theta-4}\left[f\left[(\theta-2-z)(z-1)\right]-\frac{r}{2}\left[(1+\theta-3z)f^{\prime}+r f^{\prime\prime} \right] \right].
\end{eqnarray}
In order to tackle the problem of constructing asymptotically \emph{hvLif} metrics, and taking into account the form of $V(\phi)$ and $Z(\phi)$ for our \emph{axion-free} model (and others present in the literature), we can start by considering these functions to have the generic form
\begin{eqnarray}
\label{eq:Vc_2}
V(\phi) &=& c_1 e^{-s_1 \phi} + c_2 e^{s_2\phi}+c_3,\\
\label{eq:Zd2}
Z(\phi) &=& d_1 e^{-t_1 \phi} + d_2 e^{t_2\phi}+d_3.\,
\end{eqnarray}

\noindent
The form of $V(\phi)$ is motivated by the expression of $V_{\rm fi}$ appearing  in the \emph{axion-free} $t^3$ model, as well as in other String Theory truncations present in the literature (see, e.g. \cite{Cvetic:1999xp}, \cite{Gouteraux:2011ce}). On the other hand, additional terms to the single-exponential gauge coupling have been introduced to mimic the quantum corrections appearing from String Theory (see, e.g. \cite{Harrison:2012vy}), in an attempt to cure the logarithmic behavior of the dilaton, which blows up in the deep IR, pointing out the non-negligibility of quantum corrections in this regime. The expressions for $V(\phi)$ and $Z(\phi)$ can be introduced in Eqs. (\ref{eqV1}) and (\ref{Zpq}), (\ref{Zp}) or (\ref{Zq}) (depending on whether we are searching for electric, magnetic or dyonic solutions) using Eq. (\ref{eqZ1}). Once this is done, we are left with two second-order differential equations for $f(r)$ which can in general be converted into a first order equation plus a constraint that remains to be fulfilled. Obtaining the general solution in the presence of so many arbitrary parameters ($c_1$, $c_2$, $c_3$, $d_1$, $d_2$, $d_3$, $s_1$, $s_2$, $t_1$, $t_2$, $z$ and $\theta$) seems not to be possible and therefore we are forced to consider further simplifications, keeping in mind that the procedure does work for other set-ups in which $Z(\phi)$ and $V(\phi)$ are given by a different choice of the parameters in (\ref{eq:Vc_2}) and (\ref{eq:Zd2}). Taking into account the form of the potentials obtained in the \emph{axion-free} $t^3$ model, let's assume $s_1=s_2$, $d_2=d_3=0$ (we allow $t_1$ to be positive or negative)

\begin{eqnarray}
\label{Vzeta}
V(\phi)&=&c_1 e^{-s_1 \phi}+c_2 e^{s_1 \phi}+c_3,\\
Z(\phi)&=& d_1 e^{-t_1 \phi}.
\end{eqnarray}

\noindent
The general form of the blackening factor, valid in all cases (electric, magnetic and dyonic), reads

\begin{equation}
{f(r)=\frac{c_3r^{\theta}}{D_3}+\frac{c_2 r^{\theta+s_1\Delta}}{D_2}+\frac{c_1 r^{\theta-s_1\Delta}}{D_1}+\frac{d_1p^2 r^{4-\theta-t_1\Delta}}{2D_p}+\frac{q^2 r^{4-\theta+t_1\Delta}}{2d_1 D_q}+K r^{2-\theta+z}}\, ,
\end{equation}

\noindent
where $ \Delta= \sqrt{(\theta-2)(\theta-2z+2)}$, $K$ is an integration constant, and

\begin{eqnarray}
D_1&=&(\theta-2)(2-2\theta+s_1\Delta+z),\\
D_2&=&(\theta-2)(2-2\theta-s_1\Delta+z),\\
D_3&=&(\theta-2)(2-2\theta+z),\\
D_p&=&(\theta-2)(2-t_1\Delta-z),\\
D_q&=&(\theta-2)(2+t_1\Delta-z).
\end{eqnarray}

\noindent
As we said, there is an additional (non trivial) constraint to be satisfied

\begin{equation}
\label{constraint}
{f^{\prime\prime}(r)-2 r^{\theta-2}\left[-c_3-c_1 r^{-s1 \Delta}-c_2 r^{s_1\Delta}-\frac{1}{2}r^{-\theta}(\theta-z-1)\left[2(-2+\theta-z)f(r)+3rf^{\prime}(r) \right] \right]=0}.
\end{equation}

\noindent
At this point, there are several ways to construct solutions. On the one hand, it is possible to impose values to $z$ and $\theta$ and find the corresponding potentials admitting solutions for particular blackening factors. On the other hand, it is possible to fix the coefficients in the exponents of $Z$ and $V$ and find the blackening factors allowed by Eq. (\ref{constraint}). We will proceed along the lines of the second possibility, looking for solutions embedded in the Supergravity $t^3$ model. Before doing so, let's consider the general case in which the exponents in $Z(\phi)$ and $V(\phi)$ are such that $s_1=\theta/\Delta$, $t_1=(4-\theta)/\Delta$, and $c_2=q=0$. The result is a family of solutions for arbitrary values of $z$ and $\theta$ determined by

\begin{equation}
c_1=\frac{d_1p^2(\theta-z-1)}{2(1-z)},
\end{equation}
\begin{equation}
f(r)=\frac{d_1p^2}{2(1-z)(z-\theta+2)}\left[1-K r^{2+z-\theta}\right],
\end{equation}

\noindent
which is well known (see, e.g.  \cite{Dong:2012se}, \cite{Gouteraux:2011ce}, \cite{Charmousis:2010zz})

\begin{equation}
\label{kach}
f(r)\sim 1-K r^{2+z-\theta}.
\end{equation}

\noindent
The same family can also be found for electric solutions setting $s_1=\theta/\Delta$, $t_1=(\theta-4)/\Delta$, and $c_1=p=0$. In that case, the solution is given by

\begin{equation}
c_1=\frac{q^2(\theta-1-z)}{2d_1(1-z)},
\end{equation}
\begin{equation}
f(r)=\frac{q^2}{2d_1(1-z)(z-\theta+2)}\left[1-K r^{2+z-\theta}\right].
\end{equation}


\subsubsection*{$t^3$-model}


\begin{enumerate}

\item  {\bf Magnetic solutions.} As we saw, a consistent truncation of the $t^3$-model can be embedded in the E.M.D. system for $s_1=1/\sqrt{3}$, $t_1=\sqrt{3}$, $c_3=0$. It turns out that setting $q=0$, it is possible to construct two families of solutions which, in the apropriate cases, asymptote to the purely \emph{hvLif} ones constructed in the previous subsection. The first one is determined by

\begin{equation}
c_1=0,~\theta=2\left(1-\frac{\Delta}{\sqrt{3}}\right),~c_2=Ap^2,
\end{equation}

\noindent
where $A$ is a constant depending on $z$ and $\theta$. The blackening factor is given by

\begin{equation}
f(r)=Cp^2r^{\left(2-\frac{\Delta}{\sqrt{3}}\right)}+K r^{\left(\frac{2\Delta}{\sqrt{3}}+z\right)},
\end{equation}

\noindent
where $C$ is another $z,~\theta$-dependent constant. Needless to say, the metric will not, in general, asymptote to a \emph{hvLif} (with exponents $z,~\theta$) as $r\rightarrow 0$ except for particular values of $\theta$ and $z$. However, if we choose ${\theta=-2,~z=3/2}$, $c_2=-9p^2$, we find

\begin{equation}
f(r)=\frac{4p^2}{11}\left[1-K r^{\frac{11}{2}} \right].
\end{equation}

\noindent
The second family is characterized by

\begin{equation}
c_2=0,~\theta=\left(2-\frac{\Delta}{\sqrt{3}}\right),~c1=Ap^2,
\end{equation}

\noindent
where $A$ is another constant, and the blackening factor reads

\begin{equation}
f(r)=C p^2 r^{\left(2-\frac{2\Delta}{\sqrt{3}}\right)}+K r^{\left(\frac{\Delta}{\sqrt{3}}+z\right)}.
\end{equation}

\noindent
If we set ${\theta=1,~z=3}$, it becomes

\begin{equation}
f(r)=\frac{p^2}{8}\left[1-K r^4 \right]
\end{equation}

\noindent
which, as we will see in a moment, is a particular a case of a dyonic solution admitted by the model.

\item {\bf Electric solutions.} Similarly, we can construct two families of electric solutions. The first one is characterized by

\begin{equation}
c_1=0,~\theta=\left(2+\frac{\Delta}{\sqrt{3}}\right),~c_2=Aq^2,
\end{equation}

\noindent
where, once more, $A$ is a constant depending on $z$ and $\theta$. The blackening factor is given by

\begin{equation}
f(r)=Cq^2r^{\left(2+\frac{2\Delta}{\sqrt{3}}\right)}+K r^{\left(-\frac{\Delta}{\sqrt{3}}+z\right)},
\end{equation}

whereas for the second

\begin{equation}
c_1=Aq^2,~\theta=2\left(1+\frac{\Delta}{\sqrt{3}}\right),~c_2=0,
\end{equation}

\begin{equation}
f(r)=Cq^2r^{\left(2+\frac{\Delta}{\sqrt{3}}\right)}+K r^{\left(-\frac{2\Delta}{\sqrt{3}}+z\right)}.
\end{equation}

In contradistinction to the magnetic cases, for no values of $(\theta,z)$ the above solutions take the form of Eq. (\ref{kach}). This is obviously connected to the fact that no purely \emph{hvLif} electric solutions exist in this model for non-constant dilaton and scalar potential, as we saw before.

\noindent
\item {\bf Dyonic solutions.} It is possible to show that a dyonic solution does exist for $\theta=1$, $z=3$, $c_2=0$, and $c_1=-3p^2/2$, with a blackening factor given by

\begin{equation}
f(r)=\frac{p^2}{8}\left[1-Kr^4+\frac{q^2}{p^2}r^6\right].
\end{equation}

\noindent
The corresponding metric Eq.(\ref{eq:ABmetric}) reads (after the redefinitions $dR^2=8dr^2/p^2$, $dT^2=8dt^2/p^2$)

\begin{equation}
{ds^2_f=\frac{L^2}{R}\left\{\left[1-K R^4+\frac{p^4 q^2}{512} R^6 \right]\frac{dT^2}{R^4} -\frac{dR^2}{\left[1-K R^4+\frac{p^4 q^2}{512} R^6 \right]} -d\vec{x}^2\right\}}.
\end{equation}

\noindent
It asymptotes to a \emph{hvLif} as $R\rightarrow 0$ with $\theta=1,~ z=3$, and to a different one as $R\rightarrow \infty$ with $\theta=5/2$, $z=3/2$ as it can be seen by taking the limit in the previous expression, and defining $\rho \sim R^{-2}$

\begin{equation}
ds^2_f\overset{R\rightarrow +\infty}{\sim}\frac{L^2}{R}\left[R^2 dT^2 - \frac{dR^2}{R^6} -d\vec{x}^2\right],
\end{equation}

\begin{equation}
ds^2_f \overset{[R\rightarrow +\infty,~R^{-2}=\rho]}{\sim} L^2\rho^{1/2}\left[\frac{dT^2}{\rho} - d\rho^2 -d\vec{x}^2\right],
\end{equation}

\noindent
which corresponds to $\theta=5/2,~z=3/2$. The value of $K$ can be fixed in a way such that $\exists~R_h\in  \mathbb{R}^{+}~/~f(R_h)=0$, or chosen to get a positive-definite metric in the whole spacetime.\\

\end{enumerate}

In the previous section, we constructed two consistent truncations of this model (which we called "I" and "II"). The first one is such that $c_2=0 \Rightarrow c_1=0$, and hence the solution can be embedded in that model only for a vanishing $V_{\rm fi}$ and magnetic charge. For the second, in turn, we get the conditions $g_0=0$, $\left( g^1\right)^2=p^2/2$. It is interesting to investigate how the solution gets modified by turning off the electric or the magnetic charge. Obviously, setting $q=0$ does not change the $R \rightarrow 0$ behavior, but does change the $R \rightarrow +\infty$ one. In such a case, the metric becomes

\begin{equation}
ds^2_f\overset{[R\rightarrow +\infty, R^{-1}=\rho]}{\sim}\rho\left[dT^2-d\rho^2-d\vec{x}^2 \right],
\end{equation}

\noindent
which is conformal to Minkowski, and corresponds to a \emph{hvLif} with $\theta=3,~ z=1$. On the other hand, restoring $q$ and setting $p=0$, imposes the vanishing of $V_{\rm fi}$, and the solution is $\theta=3,~z=1$ as $R\rightarrow 0$, and again $\theta=5/2,~ z=3/2$ as $R\rightarrow + \infty$.  \\

\noindent
It turns out that there exists another dyonic solution for ${\theta=5/2,~z=3/2}$\footnote{One may wonder why we did not find a purely \emph{hvLif} for these values of the exponents in the previous subsection. The reason is that for $\theta=5/2,~z=3/2$ we have $\mathcal{X}_{(\theta,z)}=0$, which implies the vanishing of $V_{\rm fi}$ in the purely \emph{hvLif} case. In fact, to recover the pure solution, we have to set $K=q=c_2=0$, and since we have already set $c_1=0$, this would make $V_{\rm fi}=0$.}. This is somehow "dual" to the previous one, as it presents the same IR and UV behavior but with both regimes interchanged. It is characterized by $c_1=0$, $c_2=-\frac{3q^2}{8}$, and

\begin{equation}
f(r)=2p^2\left[1-Kr+\frac{q^2}{16p^2}r^3\right].
\end{equation}

\noindent
In our "I" truncation, $c^I_2=-g_1^2/3\Rightarrow g_1^2=9q^2/8$. Making the redefinitions $dR^2=dr^2/(2p^2),~dT^2=\sqrt{2}p dt^2$, it reads

\begin{equation}
{ds^2_f=L^2R^{1/2}\left\{\left[1-KR+\frac{pq^2}{4\sqrt{2}}R^3 \right] \frac{dT^2}{R} -\frac{dR^2}{\left[1-KR+\frac{pq^2}{4\sqrt{2}}R^3 \right]}-d\vec{x}^2\right\}}.
\end{equation}

\noindent
As $R\rightarrow +\infty$, this becomes

\begin{equation}
ds^2_f\overset{[UV,~R=\rho^{-2}]}{\sim} \frac{L^2 }{\rho}\left[\frac{dT^2}{\rho^4}-d\rho^2-d\vec{x}^2 \right],
\end{equation}

\noindent
up to constants, which corresponds to a \emph{hvLif} with $\theta=1,~z=3$.

\subsubsection{Constant scalar field}

Let's consider now the case of a constant scalar field, $\phi^{\prime}=0$. As explained in section \ref{sec:effectiveaction}, we consider

\begin{equation}
\label{vzphi}
\partial_{\phi}V_{\rm bh}=\partial_{\phi}V=0.
\end{equation}

\noindent
In this case, the potential and the coupling become constant and we can write $V=\Lambda$, $Z=-Z_0^2$. When $Z$ and $V$ are given by Eqs. (\ref{eq:Zd2}) and (\ref{eq:Vc_2}), Eq. (\ref{vzphi}) translates into

\begin{equation}
\partial_{\phi}V_{\rm bh}(\phi=0)=\partial_{\phi}Z\left(p^2-\frac{q^2}{Z^2} \right)|_{\phi=0}=(-t_1d_1+t_2d_2)\left(p^2-\frac{q^2}{(d_1+d_2)^2} \right)=0,
\end{equation}
\begin{equation}
\partial_{\phi}V(\phi=0)=(s_2 c_2-s_1 c_1)=0,
\end{equation}

\noindent
where we have imposed $\phi=0$ to be a critical point of the potentials. We choose to fulfill the first condition demanding $(d_1+d_2)^2=q^2/p^2$ which, when $d_1=0$, reads $d_2=-|q/p|$. On the other hand, the second condition is $s_1 c_1=s_2c_2$, that becomes $c_1=c_2$ when both exponents ($s_1 $ and $s_2$) coincide. After imposing these constraints, $V$ and $Z$ become
\begin{eqnarray}
V&=& c_2\left(\frac{s_2}{s_1}+1 \right)+c_3\equiv\Lambda(=2 c_2+c_3~ \text{if}~ s_2=s_1),\\
Z&=& - \left|\frac{q}{p}\right|\equiv -Z_0^2.
\end{eqnarray}

\noindent
We have two cases: $z=1+\theta/2$ and $\theta=2$ (and the one in the intersection: $z=2$, $\theta=2$).\\

\begin{enumerate}

\item ${z=1+\frac{\theta}{2},~\theta\neq 2}$. In this situation, it is possible to find a solution which imposes no further constraints on $V$ and $V_{\rm bh}$. This reads
    \begin{equation}
\label{z1t2}
    f(r)=-K r^{3-\theta/2}+\frac{\left[12Z_0^2 r^{4-\theta}-2\Lambda r^{\theta} \right]}{3(\theta-2)^2}.
    \end{equation}
    The case \makebox{$z=1,\theta=0$}, in which we expect to recover $aDS_4$ asymptotically is a particularization of this. The blackening factor reads then
    \begin{equation}
    f(r)=-\frac{\Lambda}{6}-K r^3+Z_0^2 r^4,
    \end{equation}
    Assuming a negative cosmological constant, $\Lambda=-|\Lambda|$, this can be rewritten as
    \begin{equation}
    f(r)=\frac{|\Lambda|}{6}\left[1-K r^3+\frac{6Z_0^2}{|\Lambda|}r^4 \right].
    \end{equation}
    If we define $dT^2=|\Lambda|dt^2/6$, $dR^2=6dr^2/|\Lambda|$, the metric Eq. (\ref{eq:ABmetric}) becomes
    \begin{equation}
    \displaystyle ds_f^2=\frac{L^2}{R^2}\left\{\left[1-K R^3+\frac{|\Lambda|Z_0^2}{6} R^4 \right]dT^2-\frac{dR^2}{\left[1-K R^3+\frac{|\Lambda|Z_0^2}{6} R^4 \right]}-d\vec{x}^2\right\},
    \end{equation}
    which, of course, asymptotes to $aDS_4$ as $R\rightarrow 0$, and is such that $\exists~R_h\in  \mathbb{R}^{+}~/~f(R_h)=0$ for $K>0$. Similarly, the metric blows up as $R\rightarrow \infty$, behaving as a \emph{hvLif} with $\theta=4,~z=3$. Indeed,

    \begin{equation}
    ds_f^2\overset{R\rightarrow \infty}{\sim}\frac{L^2}{R^2}\left[R^4 dT^2-\frac{dR^2}{R^4}-d\vec{x}^2\right]
    \end{equation}

    \noindent
    up to constants; if we make now the change $\rho \sim 1/R$

    \begin{equation}
        ds_f^2\overset{\rho\rightarrow 0}{\sim}L^{\prime 2}\rho^2\left[\frac{dT^2}{\rho^4}-d\rho^2-d\vec{x}^2\right],
    \end{equation}

    \noindent
    we find a \emph{hvLif} metric with $\theta=4,~z=3$ as we have said. If we plug these values $\theta=4,~z=3$ in Eq. (\ref{z1t2}) we find a new solution, which behaves asymptotically as this one (with the $IR$ and $UV$ regions interchanged). Indeed, its blackening factor reads

\begin{equation}
f(r)=Z_0^2\left[1-K r +\frac{|\Lambda|}{6Z_0^2}r^4 \right],
\end{equation}

\noindent
and with the redefinitions $dR^2=dr^2/Z_0^2,~dT^2=dt^2/Z_0^2$

\begin{equation}
ds_f^2=L^2 R^2\left\{ \left[1-K R+\frac{|\Lambda|Z_0^2}{6}R^4 \right]\frac{dT^2}{R^4}-\frac{dR^2}{\left[1-K R+\frac{|\Lambda|Z_0^2}{6}R^4 \right]}-d\vec{x}^2 \right\} \, .
\end{equation}

\noindent
As $R\rightarrow 0$, it becomes a \emph{hvLif} with $\theta=4,~z=3$, and as $R\rightarrow \infty$,

\begin{equation}
ds^2_f=L^2 R^2 \left[dT^2-\frac{dR^2}{R^4}-d\vec{x}^2 \right],
\end{equation}

\noindent
which we can rewrite as ($\rho=1/R$)

\begin{equation}
ds^2_f=\frac{{L^{\prime}}^2 }{\rho^2} \left[dT^2-d\rho^2-d\vec{x}^2 \right],
\end{equation}

\noindent
which is $aDS_4$.\\

\item ${\theta=2}$. This case imposes the constraint $Z_0^2=-\frac{\Lambda}{2}$, and can be solved for any value of $z$. The general form of $f(r)$, which applies for $z\neq 2$ is now
\begin{equation}
f(r)=\frac{2Z_0^2 r^2}{(z-2)^2}+r^{z}K_1+r^{2(z-1)}K_2,
\end{equation}
whereas for $z=2$ we have
\begin{equation}
\label{flog}
f(r)=2 r^2\log(r) \left[K_2+Z_0^2 \log(r) \right]+K_1 r^2.
\end{equation}

\noindent
For example, if we consider the case $\theta=2$, $z=1$, we inmediatly find the asymptotically flat metric (as $r\rightarrow 0$)

\begin{equation}
f(r)=1-Kr+2Z_0^2r^2,
\end{equation}
\begin{equation}
\label{theta2z1}
ds^2_f=l^2\left\{ dt^2\left[1-Kr+2Z_0^2r^2 \right]-\frac{dr^2}{\left[1-Kr+2Z_0^2 r^2 \right]}-d\vec{x}^2 \right\} \, ,
\end{equation}

\noindent
for which, once more $\exists~r_h\in  \mathbb{R}^{+}~/~f(r_h)=0$ for $K>0$. As $r\rightarrow\infty$, up to constants, it behaves as

\begin{equation}
ds^2_f\overset{[R\rightarrow +\infty]}{\sim}{l^{\prime}}^2 \left[e^{2R}dt^2-dR^2-d\vec{x}^2\right]\, ,
\end{equation}

\noindent
where we defined $R=\log r$. This is nothing but $aDS_2\times \mathbb{R}_2$. On the other hand, if we set $\theta=2$, $z=2$, from Eq. (\ref{flog}) we find

\begin{equation}
f(r)=2 r^2\log(r) \left[-K+Z_0^2 \log(r) \right]+r^2\overset{[R=\log r]}{=}e^{2R}\left[1-KR+2Z_0^2R^2 \right],
\end{equation}
\begin{equation}
ds^2_f=l^2\left\{ dt^2\left[1-KR+2Z_0^2R^2 \right]-\frac{dR^2}{\left[1-KR+2Z_0^2R^2 \right]}-d\vec{x}^2  \right\}
\end{equation}

\noindent
which is nothing but Eq. (\ref{theta2z1}).

\end{enumerate}


\section{Conclusions}
\label{sec:conclusions}

We have studied purely \emph{hvLif} and \emph{hvLif}-like solutions of the general class of theories defined by the Lagrangian (\ref{eq:action}), which covers any theory of gravity coupled to an arbitrary number of scalars and vector fields up to two derivatives. We have obtained the general effective one-dimensional equations of motion that need to be solved in order to obtain \emph{hvLif}-like solutions. \\

The general analysis is intended to complete the  case-by-case results present in the literature in a unified framework: given a particular kinetic matrix $\left(I_{\Lambda\Sigma}(\phi),~R_{\Lambda\Sigma}(\phi)\right)$, a scalar metric $\mathcal{G}_{ij}(\phi)$ and a scalar potential $V(\phi)$, the equations of motion of the theory follow trivially by plugging them into (\ref{eq:1dEqU})-(\ref{eq:1dEqphi}) and the Hamiltonian constraint (\ref{eq:hamiltonian}).\\

For this broad family of theories, we have discussed the existence and properties of purely \emph{hvLif} solutions attending to the presence (or absence) of non-constant scalar fields and non-vanishing black hole and scalar potentials. \\

In the context of $\mathcal{N}=2$ FI Supergravity, we have studied the $t^3$-model, for which we have explicitly constructed two consistent \emph{axion-free} embeddings in the E.M.D. system, one of which is, in turn, embedded in Type-IIB String Theory for a particular choice of embedding tensor $\theta_{M}$. \\

In addition, we obtained the general form of the $f(r)$ function (for the set of metrics determined by Eqs. (\ref{ansatz}) and (\ref{eq:ABmetric})), up to a constraint, for a rather general family of (Supergravity inspired) scalar and black-hole potentials, and explicitly constructed some dyonic solutions for the $t^3$ truncations considered. We have provided a straightforward procedure to construct asymptotically \emph{hvLif} solutions covered by Eqs. (\ref{ansatz}) and (\ref{eq:ABmetric}) for the family of theories specified by Eqs. (\ref{Vzeta}), reducing the task to solving a single algebraic constraint given by Eq. (\ref{constraint}).\\

We have avoided, on purpose, the term \emph{black hole} to denote the \emph{hvLif}-like solutions obtained in this paper. The reason is that, although they look like black holes, a complete and rigorous proof  (for example by constructing the corresponding Penrose-Carter diagram) is still missing. Therefore, any results obtained from them implicitly assuming that they do represent a black hole must be interpreted carefully, knowing that those would be yet to be proven statements. \\

As a final remark, we would like to point out that, thanks to the \emph{BH-hvLif-Topological} triality (BHvTriality) discovered in \cite{Bueno:2012sd}, the new fascinating results that are being obtained in the context of static, spherically symmetric black holes in ungauged $\mathcal{N}=2,~d=4$ Supergravity \cite{Meessen:2011aa,Meessen:2012su,Bueno:2012jc,Galli:2012pt,Galli:2012jh,Galli:2012ji} can also be applied to \emph{hvLif}, giving therefore, the first examples of \emph{hvLif}-like solutions in the presence of quantum corrections induced by Type-IIA String Theory Calabi-Yau compatifications.\\

{\bf Note added:} During the very last stage of this project, the very interesting Ref. \cite{Cremonini:2012ir} appeared, containing a minor overlap with our work.


\section*{Acknowledgments}

The authors thank Tom\'as Ort\'in for useful disussions. WC acknowledges the hospitality at IFT where the final part of this work was performed
This work has been supported in part by the Spanish Ministry of Science and
Education grant FPA2009-07692, the Comunidad de Madrid grant HEPHACOS
S2009ESP-1473 and the Spanish Consolider-Ingenio 2010 program CPAN
CSD2007-00042. The work of P.B. and C.S.S. has been supported by the JAE-predoc grants JAEPre 2011 00452 and JAEPre 2010 00613.  P.B. wishes to thank N. Fuster for her unshakeable support. C.S.S. wishes to thank S. Ruiz for her unswerving support.



\renewcommand{\leftmark}{\MakeUppercase{Bibliography}}
\phantomsection
\bibliographystyle{JHEP}
\bibliography{References}

\providecommand{\href}[2]{#2}\begingroup\raggedright\begin{thebibliography}{10}

\bibitem{Huijse:2011ef}
L.~Huijse, S.~Sachdev, and B.~Swingle, {\it {Hidden Fermi surfaces in
  compressible states of gauge-gravity duality}},  {\em Phys.Rev.} {\bf B85}
  (2012) 035121, [\href{http://xxx.lanl.gov/abs/1112.0573}{{\tt
  arXiv:1112.0573}}].

\bibitem{Iizuka:2011hg}
N.~Iizuka, N.~Kundu, P.~Narayan, and S.~P. Trivedi, {\it {Holographic Fermi and
  Non-Fermi Liquids with Transitions in Dilaton Gravity}},  {\em JHEP} {\bf
  1201} (2012) 094, [\href{http://xxx.lanl.gov/abs/1105.1162}{{\tt
  arXiv:1105.1162}}].

\bibitem{Shaghoulian:2011aa}
E.~Shaghoulian, {\it {Holographic Entanglement Entropy and Fermi Surfaces}},
  {\em JHEP} {\bf 1205} (2012) 065,
  [\href{http://xxx.lanl.gov/abs/1112.2702}{{\tt arXiv:1112.2702}}].

\bibitem{Iizuka:2012pn}
N.~Iizuka, S.~Kachru, N.~Kundu, P.~Narayan, N.~Sircar, et~al., {\it {Extremal
  Horizons with Reduced Symmetry: Hyperscaling Violation, Stripes, and a
  Classification for the Homogeneous Case}},  {\em JHEP} {\bf 1303} (2013) 126,
  [\href{http://xxx.lanl.gov/abs/1212.1948}{{\tt arXiv:1212.1948}}].

\bibitem{Ogawa:2011bz}
N.~Ogawa, T.~Takayanagi, and T.~Ugajin, {\it {Holographic Fermi Surfaces and
  Entanglement Entropy}},  {\em JHEP} {\bf 1201} (2012) 125,
  [\href{http://xxx.lanl.gov/abs/1111.1023}{{\tt arXiv:1111.1023}}].

\bibitem{Edalati:2012tc}
M.~Edalati, J.~F. Pedraza, and W.~Tangarife~Garcia, {\it {Quantum Fluctuations
  in Holographic Theories with Hyperscaling Violation}},  {\em Phys.Rev.} {\bf
  D87} (2013) 046001, [\href{http://xxx.lanl.gov/abs/1210.6993}{{\tt
  arXiv:1210.6993}}].

\bibitem{Kim:2012pd}
B.~S. Kim, {\it {Hyperscaling violation : a unified frame for effective
  holographic theories}},  {\em JHEP} {\bf 1211} (2012) 061,
  [\href{http://xxx.lanl.gov/abs/1210.0540}{{\tt arXiv:1210.0540}}].

\bibitem{Alishahiha:2012qu}
M.~Alishahiha, E.~O~Colgain, and H.~Yavartanoo, {\it {Charged Black Branes with
  Hyperscaling Violating Factor}},  {\em JHEP} {\bf 1211} (2012) 137,
  [\href{http://xxx.lanl.gov/abs/1209.3946}{{\tt arXiv:1209.3946}}].

\bibitem{Gath:2012pg}
J.~Gath, J.~Hartong, R.~Monteiro, and N.~A. Obers, {\it {Holographic Models for
  Theories with Hyperscaling Violation}},
  \href{http://xxx.lanl.gov/abs/1212.3263}{{\tt arXiv:1212.3263}}.

\bibitem{Gouteraux:2012yr}
B.~Gouteraux and E.~Kiritsis, {\it {Quantum critical lines in holographic
  phases with (un)broken symmetry}},  {\em JHEP} {\bf 1304} (2013) 053,
  [\href{http://xxx.lanl.gov/abs/1212.2625}{{\tt arXiv:1212.2625}}].

\bibitem{Alishahiha:2012cm}
M.~Alishahiha and H.~Yavartanoo, {\it {On Holography with Hyperscaling
  Violation}},  {\em JHEP} {\bf 1211} (2012) 034,
  [\href{http://xxx.lanl.gov/abs/1208.6197}{{\tt arXiv:1208.6197}}].

\bibitem{Narayan:2012hk}
K.~Narayan, {\it {On Lifshitz scaling and hyperscaling violation in string
  theory}},  {\em Phys.Rev.} {\bf D85} (2012) 106006,
  [\href{http://xxx.lanl.gov/abs/1202.5935}{{\tt arXiv:1202.5935}}].

\bibitem{Dey:2012tg}
P.~Dey and S.~Roy, {\it {Lifshitz-like space-time from intersecting branes in
  string/M theory}},  {\em JHEP} {\bf 1206} (2012) 129,
  [\href{http://xxx.lanl.gov/abs/1203.5381}{{\tt arXiv:1203.5381}}].

\bibitem{Dey:2012rs}
P.~Dey and S.~Roy, {\it {Intersecting D-branes and Lifshitz-like space-time}},
  {\em Phys.Rev.} {\bf D86} (2012) 066009,
  [\href{http://xxx.lanl.gov/abs/1204.4858}{{\tt arXiv:1204.4858}}].

\bibitem{Charmousis:2012dw}
C.~Charmousis, B.~Gouteraux, and E.~Kiritsis, {\it {Higher-derivative
  scalar-vector-tensor theories: black holes, Galileons, singularity cloaking
  and holography}},  {\em JHEP} {\bf 1209} (2012) 011,
  [\href{http://xxx.lanl.gov/abs/1206.1499}{{\tt arXiv:1206.1499}}].

\bibitem{Ammon:2012je}
M.~Ammon, M.~Kaminski, and A.~Karch, {\it {Hyperscaling-Violation on Probe
  D-Branes}},  {\em JHEP} {\bf 1211} (2012) 028,
  [\href{http://xxx.lanl.gov/abs/1207.1726}{{\tt arXiv:1207.1726}}].

\bibitem{Bhattacharya:2012zu}
J.~Bhattacharya, S.~Cremonini, and A.~Sinkovics, {\it {On the IR completion of
  geometries with hyperscaling violation}},  {\em JHEP} {\bf 1302} (2013) 147,
  [\href{http://xxx.lanl.gov/abs/1208.1752}{{\tt arXiv:1208.1752}}].

\bibitem{Dey:2012fi}
P.~Dey and S.~Roy, {\it {Lifshitz metric with hyperscaling violation from
  NS5-Dp states in string theory}},  {\em Phys.Lett.} {\bf B720} (2013)
  419--423, [\href{http://xxx.lanl.gov/abs/1209.1049}{{\tt arXiv:1209.1049}}].

\bibitem{Kim:2012nb}
B.~S. Kim, {\it {Schr\'odinger Holography with and without Hyperscaling
  Violation}},  {\em JHEP} {\bf 1206} (2012) 116,
  [\href{http://xxx.lanl.gov/abs/1202.6062}{{\tt arXiv:1202.6062}}].

\bibitem{Perlmutter:2012he}
E.~Perlmutter, {\it {Hyperscaling violation from supergravity}},  {\em JHEP}
  {\bf 1206} (2012) 165, [\href{http://xxx.lanl.gov/abs/1205.0242}{{\tt
  arXiv:1205.0242}}].

\bibitem{Gouteraux:2011qh}
B.~Gouteraux, J.~Smolic, M.~Smolic, K.~Skenderis, and M.~Taylor, {\it
  {Holography for Einstein-Maxwell-dilaton theories from generalized
  dimensional reduction}},  {\em JHEP} {\bf 1201} (2012) 089,
  [\href{http://xxx.lanl.gov/abs/1110.2320}{{\tt arXiv:1110.2320}}].

\bibitem{Sadeghi:2012vv}
J.~Sadeghi, B.~Pourhasan, and F.~Pourasadollah, {\it {Thermodynamics of
  Schršdinger black holes with hyperscaling violation}},  {\em Phys.Lett.} {\bf
  B720} (2013) 244--249, [\href{http://xxx.lanl.gov/abs/1209.1874}{{\tt
  arXiv:1209.1874}}].

\bibitem{Donos:2012yu}
A.~Donos, J.~P. Gauntlett, J.~Sonner, and B.~Withers, {\it {Competing orders in
  M-theory: superfluids, stripes and metamagnetism}},  {\em JHEP} {\bf 1303}
  (2013) 108, [\href{http://xxx.lanl.gov/abs/1212.0871}{{\tt
  arXiv:1212.0871}}].

\bibitem{Copsey:2010ya}
K.~Copsey and R.~Mann, {\it {Pathologies in Asymptotically Lifshitz
  Spacetimes}},  {\em JHEP} {\bf 1103} (2011) 039,
  [\href{http://xxx.lanl.gov/abs/1011.3502}{{\tt arXiv:1011.3502}}].

\bibitem{Copsey:2012gw}
K.~Copsey and R.~Mann, {\it {Singularities in Hyperscaling Violating
  Spacetimes}},  {\em JHEP} {\bf 1304} (2013) 079,
  [\href{http://xxx.lanl.gov/abs/1210.1231}{{\tt arXiv:1210.1231}}].

\bibitem{Bao:2012yt}
N.~Bao, X.~Dong, S.~Harrison, and E.~Silverstein, {\it {The Benefits of Stress:
  Resolution of the Lifshitz Singularity}},
  \href{http://xxx.lanl.gov/abs/1207.0171}{{\tt arXiv:1207.0171}}.

\bibitem{Copsey:2011ek}
K.~Copsey and R.~B. Mann, {\it {Hidden Singularities and Closed Timelike Curves
  in A Proposed Dual for Lifshitz-Chern-Simons Gauge Theories}},
  \href{http://xxx.lanl.gov/abs/1112.0578}{{\tt arXiv:1112.0578}}.

\bibitem{Horowitz:2011gh}
G.~T. Horowitz and B.~Way, {\it {Lifshitz Singularities}},  {\em Phys.Rev.}
  {\bf D85} (2012) 046008, [\href{http://xxx.lanl.gov/abs/1111.1243}{{\tt
  arXiv:1111.1243}}].

\bibitem{Balasubramanian:2011ua}
K.~Balasubramanian and J.~McGreevy, {\it {String theory duals of
  Lifshitz-Chern-Simons gauge theories}},
  \href{http://xxx.lanl.gov/abs/1111.0634}{{\tt arXiv:1111.0634}}.

\bibitem{Kachru:2008yh}
S.~Kachru, X.~Liu, and M.~Mulligan, {\it {Gravity Duals of Lifshitz-like Fixed
  Points}},  {\em Phys.Rev.} {\bf D78} (2008) 106005,
  [\href{http://xxx.lanl.gov/abs/0808.1725}{{\tt arXiv:0808.1725}}].

\bibitem{Taylor:2008tg}
M.~Taylor, {\it {Non-relativistic holography}},
  \href{http://xxx.lanl.gov/abs/0812.0530}{{\tt arXiv:0812.0530}}.

\bibitem{Azeyanagi:2009pr}
T.~Azeyanagi, W.~Li, and T.~Takayanagi, {\it {On String Theory Duals of
  Lifshitz-like Fixed Points}},  {\em JHEP} {\bf 0906} (2009) 084,
  [\href{http://xxx.lanl.gov/abs/0905.0688}{{\tt arXiv:0905.0688}}].

\bibitem{Balasubramanian:2010uk}
K.~Balasubramanian and K.~Narayan, {\it {Lifshitz spacetimes from AdS null and
  cosmological solutions}},  {\em JHEP} {\bf 1008} (2010) 014,
  [\href{http://xxx.lanl.gov/abs/1005.3291}{{\tt arXiv:1005.3291}}].

\bibitem{Donos:2010tu}
A.~Donos and J.~P. Gauntlett, {\it {Lifshitz Solutions of D=10 and D=11
  supergravity}},  {\em JHEP} {\bf 1012} (2010) 002,
  [\href{http://xxx.lanl.gov/abs/1008.2062}{{\tt arXiv:1008.2062}}].

\bibitem{Gregory:2010gx}
R.~Gregory, S.~L. Parameswaran, G.~Tasinato, and I.~Zavala, {\it {Lifshitz
  solutions in supergravity and string theory}},  {\em JHEP} {\bf 1012} (2010)
  047, [\href{http://xxx.lanl.gov/abs/1009.3445}{{\tt arXiv:1009.3445}}].

\bibitem{Narayan:2011az}
K.~Narayan, {\it {Lifshitz-like systems and AdS null deformations}},  {\em
  Phys.Rev.} {\bf D84} (2011) 086001,
  [\href{http://xxx.lanl.gov/abs/1103.1279}{{\tt arXiv:1103.1279}}].

\bibitem{Chemissany:2011mb}
W.~Chemissany and J.~Hartong, {\it {From D3-Branes to Lifshitz Space-Times}},
  {\em Class.Quant.Grav.} {\bf 28} (2011) 195011,
  [\href{http://xxx.lanl.gov/abs/1105.0612}{{\tt arXiv:1105.0612}}].

\bibitem{Chemissany:2012du}
W.~Chemissany, D.~Geissbuhler, J.~Hartong, and B.~Rollier, {\it {Holographic
  Renormalization for z=2 Lifshitz Space-Times from AdS}},  {\em
  Class.Quant.Grav.} {\bf 29} (2012) 235017,
  [\href{http://xxx.lanl.gov/abs/1205.5777}{{\tt arXiv:1205.5777}}].

\bibitem{Danielsson:2009gi}
U.~H. Danielsson and L.~Thorlacius, {\it {Black holes in asymptotically
  Lifshitz spacetime}},  {\em JHEP} {\bf 0903} (2009) 070,
  [\href{http://xxx.lanl.gov/abs/0812.5088}{{\tt arXiv:0812.5088}}].

\bibitem{Bertoldi:2009vn}
G.~Bertoldi, B.~A. Burrington, and A.~Peet, {\it {Black Holes in asymptotically
  Lifshitz spacetimes with arbitrary critical exponent}},  {\em Phys.Rev.} {\bf
  D80} (2009) 126003, [\href{http://xxx.lanl.gov/abs/0905.3183}{{\tt
  arXiv:0905.3183}}].

\bibitem{Balasubramanian:2009rx}
K.~Balasubramanian and J.~McGreevy, {\it {An Analytic Lifshitz black hole}},
  {\em Phys.Rev.} {\bf D80} (2009) 104039,
  [\href{http://xxx.lanl.gov/abs/0909.0263}{{\tt arXiv:0909.0263}}].

\bibitem{Tarrio:2011de}
J.~Tarrio and S.~Vandoren, {\it {Black holes and black branes in Lifshitz
  spacetimes}},  {\em JHEP} {\bf 1109} (2011) 017,
  [\href{http://xxx.lanl.gov/abs/1105.6335}{{\tt arXiv:1105.6335}}].

\bibitem{Hartnoll:2009sz}
S.~A. Hartnoll, {\it {Lectures on holographic methods for condensed matter
  physics}},  {\em Class.Quant.Grav.} {\bf 26} (2009) 224002,
  [\href{http://xxx.lanl.gov/abs/0903.3246}{{\tt arXiv:0903.3246}}].

\bibitem{Herzog:2009xv}
C.~P. Herzog, {\it {Lectures on Holographic Superfluidity and
  Superconductivity}},  {\em J.Phys.A} {\bf A42} (2009) 343001,
  [\href{http://xxx.lanl.gov/abs/0904.1975}{{\tt arXiv:0904.1975}}].

\bibitem{McGreevy:2009xe}
J.~McGreevy, {\it {Holographic duality with a view toward many-body physics}},
  {\em Adv.High Energy Phys.} {\bf 2010} (2010) 723105,
  [\href{http://xxx.lanl.gov/abs/0909.0518}{{\tt arXiv:0909.0518}}].

\bibitem{Sachdev:2011wg}
S.~Sachdev, {\it {What can gauge-gravity duality teach us about condensed
  matter physics?}},  {\em Ann.Rev.Condensed Matter Phys.} {\bf 3} (2012)
  9--33, [\href{http://xxx.lanl.gov/abs/1108.1197}{{\tt arXiv:1108.1197}}].

\bibitem{Hartnoll:2009ns}
S.~A. Hartnoll, J.~Polchinski, E.~Silverstein, and D.~Tong, {\it {Towards
  strange metallic holography}},  {\em JHEP} {\bf 1004} (2010) 120,
  [\href{http://xxx.lanl.gov/abs/0912.1061}{{\tt arXiv:0912.1061}}].

\bibitem{Donos:2010ax}
A.~Donos, J.~P. Gauntlett, N.~Kim, and O.~Varela, {\it {Wrapped M5-branes,
  consistent truncations and AdS/CMT}},  {\em JHEP} {\bf 1012} (2010) 003,
  [\href{http://xxx.lanl.gov/abs/1009.3805}{{\tt arXiv:1009.3805}}].

\bibitem{Halmagyi:2011xh}
N.~Halmagyi, M.~Petrini, and A.~Zaffaroni, {\it {Non-Relativistic Solutions of
  N=2 Gauged Supergravity}},  {\em JHEP} {\bf 1108} (2011) 041,
  [\href{http://xxx.lanl.gov/abs/1102.5740}{{\tt arXiv:1102.5740}}].

\bibitem{Hartnoll:2012wm}
S.~A. Hartnoll and E.~Shaghoulian, {\it {Spectral weight in holographic scaling
  geometries}},  {\em JHEP} {\bf 1207} (2012) 078,
  [\href{http://xxx.lanl.gov/abs/1203.4236}{{\tt arXiv:1203.4236}}].

\bibitem{Takayanagi:2012kg}
T.~Takayanagi, {\it {Entanglement Entropy from a Holographic Viewpoint}},  {\em
  Class.Quant.Grav.} {\bf 29} (2012) 153001,
  [\href{http://xxx.lanl.gov/abs/1204.2450}{{\tt arXiv:1204.2450}}].

\bibitem{Dey:2012hf}
P.~Dey and S.~Roy, {\it {Holographic entanglement entropy of the near horizon
  1/4 BPS F-D$p$ bound states}},  {\em Phys.Rev.} {\bf D87} (2013) 066001,
  [\href{http://xxx.lanl.gov/abs/1208.1820}{{\tt arXiv:1208.1820}}].

\bibitem{Kulaxizi:2012gy}
M.~Kulaxizi, A.~Parnachev, and K.~Schalm, {\it {On Holographic Entanglement
  Entropy of Charged Matter}},  {\em JHEP} {\bf 1210} (2012) 098,
  [\href{http://xxx.lanl.gov/abs/1208.2937}{{\tt arXiv:1208.2937}}].

\bibitem{Charmousis:2010zz}
C.~Charmousis, B.~Gouteraux, B.~Kim, E.~Kiritsis, and R.~Meyer, {\it {Effective
  Holographic Theories for low-temperature condensed matter systems}},  {\em
  JHEP} {\bf 1011} (2010) 151, [\href{http://xxx.lanl.gov/abs/1005.4690}{{\tt
  arXiv:1005.4690}}].

\bibitem{Bueno:2012sd}
P.~Bueno, W.~Chemissany, P.~Meessen, T.~Ortin, and C.~Shahbazi, {\it
  {Lifshitz-like Solutions with Hyperscaling Violation in Ungauged
  Supergravity}},  {\em JHEP} {\bf 1301} (2013) 189,
  [\href{http://xxx.lanl.gov/abs/1209.4047}{{\tt arXiv:1209.4047}}].

\bibitem{Ferrara:1997tw}
S.~Ferrara, G.~W. Gibbons, and R.~Kallosh, {\it {Black holes and critical
  points in moduli space}},  {\em Nucl.Phys.} {\bf B500} (1997) 75--93,
  [\href{http://xxx.lanl.gov/abs/hep-th/9702103}{{\tt hep-th/9702103}}].

\bibitem{Galli:2011fq}
P.~Galli, T.~Ortin, J.~Perz, and C.~S. Shahbazi, {\it {Non-extremal black holes
  of N=2, d=4 supergravity}},  {\em JHEP} {\bf 1107} (2011) 041,
  [\href{http://xxx.lanl.gov/abs/1105.3311}{{\tt arXiv:1105.3311}}].

\bibitem{Meessen:2011bd}
P.~Meessen and T.~Ortin, {\it {Non-Extremal Black Holes of N=2,d=5
  Supergravity}},  {\em Phys.Lett.} {\bf B707} (2012) 178--183,
  [\href{http://xxx.lanl.gov/abs/1107.5454}{{\tt arXiv:1107.5454}}].

\bibitem{Klemm:2012yg}
D.~Klemm and O.~Vaughan, {\it {Nonextremal black holes in gauged supergravity
  and the real formulation of special geometry}},  {\em JHEP} {\bf 1301} (2013)
  053, [\href{http://xxx.lanl.gov/abs/1207.2679}{{\tt arXiv:1207.2679}}].

\bibitem{Klemm:2012vm}
D.~Klemm and O.~Vaughan, {\it {Nonextremal black holes in gauged supergravity
  and the real formulation of special geometry II}},  {\em Class.Quant.Grav.}
  {\bf 30} (2013) 065003, [\href{http://xxx.lanl.gov/abs/1211.1618}{{\tt
  arXiv:1211.1618}}].

\bibitem{Barisch:2011ui}
S.~Barisch, G.~Lopes~Cardoso, M.~Haack, S.~Nampuri, and N.~A. Obers, {\it
  {Nernst branes in gauged supergravity}},  {\em JHEP} {\bf 1111} (2011) 090,
  [\href{http://xxx.lanl.gov/abs/1108.0296}{{\tt arXiv:1108.0296}}].

\bibitem{BarischDick:2012gj}
S.~Barisch-Dick, G.~Lopes~Cardoso, M.~Haack, and S.~Nampuri, {\it {Extremal
  black brane solutions in five-dimensional gauged supergravity}},  {\em JHEP}
  {\bf 1302} (2013) 103, [\href{http://xxx.lanl.gov/abs/1211.0832}{{\tt
  arXiv:1211.0832}}].

\bibitem{deAntonioMartin:2012bi}
A.~de~Antonio~Martin, T.~Ortin, and C.~Shahbazi, {\it {The FGK formalism for
  black p-branes in d dimensions}},  {\em JHEP} {\bf 1205} (2012) 045,
  [\href{http://xxx.lanl.gov/abs/1203.0260}{{\tt arXiv:1203.0260}}].

\bibitem{Dong:2012se}
X.~Dong, S.~Harrison, S.~Kachru, G.~Torroba, and H.~Wang, {\it {Aspects of
  holography for theories with hyperscaling violation}},  {\em JHEP} {\bf 1206}
  (2012) 041, [\href{http://xxx.lanl.gov/abs/1201.1905}{{\tt
  arXiv:1201.1905}}].

\bibitem{Andrianopoli:1996cm}
L.~Andrianopoli, M.~Bertolini, A.~Ceresole, R.~D'Auria, S.~Ferrara, et~al.,
  {\it {N=2 supergravity and N=2 superYang-Mills theory on general scalar
  manifolds: Symplectic covariance, gaugings and the momentum map}},  {\em
  J.Geom.Phys.} {\bf 23} (1997) 111--189,
  [\href{http://xxx.lanl.gov/abs/hep-th/9605032}{{\tt hep-th/9605032}}].

\bibitem{Meessen:2006tu}
P.~Meessen and T.~Ortin, {\it {The Supersymmetric configurations of N=2, D=4
  supergravity coupled to vector supermultiplets}},  {\em Nucl.Phys.} {\bf
  B749} (2006) 291--324, [\href{http://xxx.lanl.gov/abs/hep-th/0603099}{{\tt
  hep-th/0603099}}].

\bibitem{Cassani:2010uw}
D.~Cassani, G.~Dall'Agata, and A.~F. Faedo, {\it {Type IIB supergravity on
  squashed Sasaki-Einstein manifolds}},  {\em JHEP} {\bf 1005} (2010) 094,
  [\href{http://xxx.lanl.gov/abs/1003.4283}{{\tt arXiv:1003.4283}}].

\bibitem{Cassani:2011sv}
D.~Cassani and A.~F. Faedo, {\it {Constructing Lifshitz solutions from AdS}},
  {\em JHEP} {\bf 1105} (2011) 013,
  [\href{http://xxx.lanl.gov/abs/1102.5344}{{\tt arXiv:1102.5344}}].

\bibitem{Liu:2010sa}
J.~T. Liu, P.~Szepietowski, and Z.~Zhao, {\it {Consistent massive truncations
  of IIB supergravity on Sasaki-Einstein manifolds}},  {\em Phys.Rev.} {\bf
  D81} (2010) 124028, [\href{http://xxx.lanl.gov/abs/1003.5374}{{\tt
  arXiv:1003.5374}}].

\bibitem{Gauntlett:2010vu}
J.~P. Gauntlett and O.~Varela, {\it {Universal Kaluza-Klein reductions of type
  IIB to N=4 supergravity in five dimensions}},  {\em JHEP} {\bf 1006} (2010)
  081, [\href{http://xxx.lanl.gov/abs/1003.5642}{{\tt arXiv:1003.5642}}].

\bibitem{Cvetic:1999xp}
M.~Cvetic, M.~Duff, P.~Hoxha, J.~T. Liu, H.~Lu, et~al., {\it {Embedding AdS
  black holes in ten-dimensions and eleven-dimensions}},  {\em Nucl.Phys.} {\bf
  B558} (1999) 96--126, [\href{http://xxx.lanl.gov/abs/hep-th/9903214}{{\tt
  hep-th/9903214}}].

\bibitem{Gouteraux:2011ce}
B.~Gouteraux and E.~Kiritsis, {\it {Generalized Holographic Quantum Criticality
  at Finite Density}},  {\em JHEP} {\bf 1112} (2011) 036,
  [\href{http://xxx.lanl.gov/abs/1107.2116}{{\tt arXiv:1107.2116}}].

\bibitem{Harrison:2012vy}
S.~Harrison, S.~Kachru, and H.~Wang, {\it {Resolving Lifshitz Horizons}},
  \href{http://xxx.lanl.gov/abs/1202.6635}{{\tt arXiv:1202.6635}}.

\bibitem{Meessen:2011aa}
P.~Meessen, T.~Ortin, J.~Perz, and C.~S. Shahbazi, {\it {H-FGK formalism for
  black-hole solutions of N=2, d=4 and d=5 supergravity}},  {\em Phys.Lett.}
  {\bf B709} (2012) 260--265, [\href{http://xxx.lanl.gov/abs/1112.3332}{{\tt
  arXiv:1112.3332}}].

\bibitem{Meessen:2012su}
P.~Meessen, T.~Ortin, J.~Perz, and C.~S. Shahbazi, {\it {Black holes and black
  strings of N=2, d=5 supergravity in the H-FGK formalism}},  {\em JHEP} {\bf
  1209} (2012) 001, [\href{http://xxx.lanl.gov/abs/1204.0507}{{\tt
  arXiv:1204.0507}}].

\bibitem{Bueno:2012jc}
P.~Bueno, R.~Davies, and C.~S. Shahbazi, {\it {Quantum Black Holes in Type-IIA
  String Theory}},  {\em JHEP} {\bf 1301} (2013) 089,
  [\href{http://xxx.lanl.gov/abs/1210.2817}{{\tt arXiv:1210.2817}}].

\bibitem{Galli:2012pt}
P.~Galli, T.~Ortin, J.~Perz, and C.~S. Shahbazi, {\it {Black hole solutions of
  N=2, d=4 supergravity with a quantum correction, in the H-FGK formalism}},
  \href{http://xxx.lanl.gov/abs/1212.0303}{{\tt arXiv:1212.0303}}.

\bibitem{Galli:2012jh}
P.~Galli, K.~Goldstein, and J.~Perz, {\it {On anharmonic stabilisation
  equations for black holes}},  {\em JHEP} {\bf 1303} (2013) 036,
  [\href{http://xxx.lanl.gov/abs/1211.7295}{{\tt arXiv:1211.7295}}].

\bibitem{Galli:2012ji}
P.~Galli, P.~Meessen, and T.~Ortin, {\it {The Freudenthal gauge symmetry of the
  black holes of N=2,d=4 supergravity}},
  \href{http://xxx.lanl.gov/abs/1211.7296}{{\tt arXiv:1211.7296}}.

\bibitem{Cremonini:2012ir}
S.~Cremonini and A.~Sinkovics, {\it {Spatially Modulated Instabilities of
  Geometries with Hyperscaling Violation}},
  \href{http://xxx.lanl.gov/abs/1212.4172}{{\tt arXiv:1212.4172}}.

\end{thebibliography}\endgroup
\label{biblio}

\end{document}